\documentclass[fleqn,usenatbib]{mnras}

\usepackage{newtxtext,newtxmath}
\usepackage{hyperref}

\usepackage[T1]{fontenc}
\usepackage{fontawesome}

\DeclareRobustCommand{\VAN}[3]{#2}
\let\VANthebibliography\thebibliography
\def\thebibliography{\DeclareRobustCommand{\VAN}[3]{##3}\VANthebibliography}

\DeclareRobustCommand{\DA}[3]{#2}
\let\DAthebibliography\thebibliography
\def\thebibliography{\DeclareRobustCommand{\DA}[3]{##3}\DAthebibliography}

\usepackage{graphicx}	
\usepackage{amsmath}	
\usepackage{wasysym}

\newcommand{\pyaneti}{\href{https://github.com/oscaribv/pyaneti}{\texttt{pyaneti}\,\faGithub}}

\newcommand{\sshk}{$S_{\rm HK}$}
\newcommand{\lbe}{$\lambda_{\rm e}$}
\newcommand{\lbp}{$\lambda_{\rm p}$}
\newcommand{\pgp}{$P_{\rm GP}$}
\newcommand{\gcm}{${\rm g\,cm^{-3}}$}
\newcommand{\ms}{${\rm m\,s^{-1}}$}
\newcommand{\kms}{${\rm km\,s^{-1}}$}

\newcommand{\citlalicue}{\texttt{citlalicue}}
\newcommand{\citlalatonac}{\texttt{citlalatonac}}
\newcommand{\vsini}{$v \sin i$}
\newcommand{\logg}{$\log g$}
\newcommand{\ktwo}{\emph{K2}}
\newcommand{\halpha}{$\mathrm{H}_{\alpha}$}

\newcommand{\msun}{$M_{\odot}$}

\newcommand{\mearth}{$M_{\oplus}$}

\newcommand{\mjup}{$M_{\rm J}$}
\newcommand{\rjup}{$R_{\rm J}$}

\newcommand{\tess}{\emph{TESS}}

\newcommand{\jwst}{\emph{JWST}}

\newcommand{\target}{TOI-837}
\newcommand{\targetb}{TOI-837\,b}

\newcommand{\oscar}[1]{\textcolor{black}{#1}}

\newcommand{\Tzerob}[1][days]{ $ 2356.81398 \pm 0.00025 $~#1 } 
\newcommand{\Pb}[1][days]{ $ 8.3249113 \pm 0.0000036 $~#1 }

\newcommand{\bb}[1][ ]{ $ 0.920_{-0.008}^{+0.011} $~#1 } 
 
\newcommand{\rrb}[1][ ]{ $ 0.0798_{-0.0022}^{+0.0032} $~#1 } 
\newcommand{\kb}[1][${\rm m\,s^{-1}}$]{ $ 34.7_{-5.6}^{+5.3} $~#1 } 
\newcommand{\mpb}[1][$M_\mathrm{J}$]{ $ 0.379_{-0.061}^{+0.058} $~#1 } 
\newcommand{\rpb}[1][$R_\mathrm{J}$]{ $ 0.818_{-0.024}^{+0.034} $~#1 } 
 
\newcommand{\ib}[1][deg]{ $ 87.097_{-0.129}^{+0.099} $~#1 } 
\newcommand{\arb}[1][ ]{ $ 18.16_{-0.58}^{+0.50} $~#1 } 
\newcommand{\ab}[1][AU]{ $ 0.0888_{-0.0029}^{+0.0026} $~#1 } 
\newcommand{\insolationb}[1][${\rm F_{\oplus}}$]{ $ 164_{-12}^{+14} $~#1 } 
 
\newcommand{\denstrb}[1][${\rm g\,cm^{-3}}$]{ $ 1.63_{-0.15}^{+0.14} $~#1 } 
 
\newcommand{\Teqb}[1][K]{ $ 995.3_{-19.1}^{+20.4} $~#1 } 
\newcommand{\ttotb}[1][hours]{ $ 1.983_{-0.035}^{+0.034} $~#1 } 
\newcommand{\denpb}[1][${\rm g\,cm^{-3}}$]{ $ 0.89_{-0.18}^{+0.20} $~#1 } 
\newcommand{\grapb}[1][${\rm cm\,s^{-2}}$]{ $ 1603_{-340}^{+377} $~#1 } 
\newcommand{\grapparsb}[1][${\rm cm\,s^{-2}}$]{ $ 1457_{-278}^{+302} $~#1 } 
 
\newcommand{\qoneastep}[1][]{ $ 0.316_{-0.088}^{+0.115} $~#1 } 
\newcommand{\qtwoastep}[1][]{ $ 0.70_{-0.29}^{+0.19} $~#1 }

\newcommand{\HARPS}[1][${\rm km\,s^{-1}}$]{ $ -0.002_{-0.073}^{+0.076} $~#1 } 
\newcommand{\jHARPS}[1][${\rm m\,s^{-1}}$]{ $ 26.6_{-3.3}^{+3.7} $~#1 } 
\newcommand{\qoneS}[1][]{ $ 0.875_{-0.173}^{+0.091} $~#1 } 
\newcommand{\qtwoS}[1][]{ $ 0.120_{-0.086}^{+0.146} $~#1 }

\newcommand{\qoneelsaucer}[1][]{ $ 0.44_{-0.14}^{+0.18} $~#1 } 
\newcommand{\qtwoelsaucer}[1][]{ $ 0.41_{-0.28}^{+0.30} $~#1 }

\newcommand{\qoneelsaucei}[1][]{ $ 0.53_{-0.21}^{+0.24} $~#1 } 
\newcommand{\qtwoelsaucei}[1][]{ $ 0.28_{-0.20}^{+0.31} $~#1 }

\newcommand{\qoneelsauceb}[1][]{ $ 0.42_{-0.19}^{+0.24} $~#1 } 
\newcommand{\qtwoelsauceb}[1][]{ $ 0.35_{-0.24}^{+0.33} $~#1 }

\newcommand{\jtrtessm}[1][ppm]{ $ 205.4_{-96.5}^{+75.4} $~#1 }
\newcommand{\jtrtesss}[1][ppm]{ $ 248_{-145}^{+118} $~#1 }
\newcommand{\jtrelsaucer}[1][ppm]{ $ 1738_{-110}^{+114} $~#1 } 
\newcommand{\jtrelsaucei}[1][ppm]{ $ 1581_{-192}^{+191} $~#1 } 
\newcommand{\jtrelsauceb}[1][ppm]{ $ 731_{-490}^{+478} $~#1 } 
\newcommand{\jtrastep}[1][ppm]{ $ 1311_{-62}^{+63} $~#1 } 
\newcommand{\AGP}[1][]{ $ 138_{-45}^{+89} $~#1 } \newcommand{\jlambdae}[1][]{ $ 65_{-18}^{+22} $~#1 } 
\newcommand{\jlambdap}[1][]{ $ 0.40_{-0.12}^{+0.19} $~#1 } 
\newcommand{\jPGP}[1][]{ $ 2.998_{-0.010}^{+0.009} $~#1 }

\defcitealias{Bouma2020}{B20}

\title[TOI-837\,b has a 70 Earth-Mass Core]{TOI-837\,b is a Young Saturn-sized Exoplanet  with a Massive 70\,$M_{\oplus}$ Core}

\author[Barragán et al.]{
Oscar~Barragán$^{1}$\thanks{\href{mailto:oscar.barrragan@physics.ox.ac.uk}{oscar.barrragan@physics.ox.ac.uk}}, 
Haochuan~Yu$^{1}$,
Alix~Violet~Freckelton$^{2}$,
Annabella~Meech$^{3}$,
Michael~Cretignier$^{1}$, \newauthor
Annelies~Mortier$^{2}$,
Suzanne~Aigrain$^{1}$,
Baptiste~Klein$^{1}$,
Niamh K. O'Sullivan$^{1}$, 
Edward~Gillen$^{4}$,
\newauthor
Louise~Dyregaard~Nielsen$^{5}$,
Manuel~Mallorqu\'in$^{6,7}$,
Norbert Zicher$^{1}$,
\\ 
$^{1}$ Sub-department of Astrophysics, Department of Physics, University of Oxford, Oxford, OX1 3RH, UK  \label{oxford} \\
$^{2}$ School of Physics and Astronomy, University of Birmingham, Edgbaston, Birmingham B15 2TT, UK \\
$^{3}$ Center for Astrophysics, Harvard \& Smithsonian, 60 Garden St, Cambridge, MA 02138, US \\
$^{4}$ Astronomy Unit, Queen Mary University of London, Mile End Road, London E14NS, UK \\
$^{5}$ University Observatory Munich, Ludwig Maximilian University, Scheinerstrasse 1, Munich D-81679, Germany \\
$^{6}$ {Instituto de Astrof\'isica de Canarias (IAC), Calle V\'ia L\'actea s/n, 38205 La Laguna, Tenerife, Spain} \\
$^{7}$ {Departamento de Astrof\'isica, Universidad de La Laguna (ULL), 38206 La Laguna, Tenerife, Spain} \\
}

\date{Accepted XXX. Received YYY; in original form ZZZ}

\pubyear{2024}

\begin{document}
\label{firstpage}
\pagerange{\pageref{firstpage}--\pageref{lastpage}}
\maketitle

\begin{abstract}
We present an exhaustive photometric and spectroscopic analysis of TOI-837, a F9/G0 35 Myr young star, hosting a transiting exoplanet, TOI-837\,b, with an orbital period of $\sim8.32$\,d. Utilising data from \emph{TESS} and ground-based observations, we determine a planetary radius of \oscar{\rpb} for TOI-837\,b. Through detailed HARPS spectroscopic time series analysis, we derive a Doppler semi-amplitude of \oscar{\kb}, corresponding to a planetary mass of \oscar{\mpb}. The derived planetary properties suggest a substantial core of approximately 70\,$M_{\oplus}$, constituting about 60\% of the planet's total mass. This finding poses a significant challenge to existing theoretical models of core formation. We propose that future atmospheric observations with \emph{JWST} could provide insights into resolving ambiguities of TOI-837\,b, offering new perspectives on its composition, formation, and evolution.
\end{abstract}

\begin{keywords}
Planets and satellites: individual: \target\ -- Stars: activity -- Techniques: radial velocities -- Techniques: photometric
\end{keywords}



\section{Introduction}

Characterising the properties of young exoplanets ($<1$\,Gyr) at different stages is crucial to understanding the evolution and populations of exoplanets.
These planets are often elusive to detection using indirect techniques like transit and radial velocity (RV) methods, as strong stellar activity generates stellar signals that often overshadows    their signals in photometric and spectroscopic data. 
Recently, tens of young transiting exoplanets have been discovered  \citep[e.g.,][]{Bouma2020,David2018,Hobson2021,Martioli2021,Mann2021,Rizutto2020,Barragan2022}. This success is largely attributed to missions like \ktwo\  \citep[][]{Howell2014} and the Transiting Exoplanet Survey Satellite \citep[\tess;][]{Ricker2015} that provide thousands of light curves of young stars.

After identifying a young transiting exoplanet, the next step typically involves spectroscopic observations to observe the star's RV variations caused by the planet to measure its mass.
However, separating signals from the planet and star in radial velocity data remains challenging.
Gaussian Processes (GPs) have emerged as a favoured method for modelling activity-induced radial velocities, offering a flexible way to represent stochastic variations, such as the quasi-periodic stellar signals \citep[see e.g.,][]{Aigrain2022}. 
Multiple authors have used GPs, and variations of them such as multidimensional GPs \citep[see e.g.,][]{Rajpaul2015,pyaneti2}, to detect RV planetary signals on spectroscopic time series of active young stars \citep[e.g.,][]{Barragan2019,Barragan2023,Mantovan2023,Mallorquin2023,Zicher2022}.
Nevertheless, although widely embraced by the community, there exists a notable concern that the GP activity model may inadvertently mask or alter potential planetary signals \citep[see e.g., discussions in][]{Ahrer2021,Rajpaul2021}. A recent example of this comes from \citet{Suarez2022}. The authors, utilising GPs, discovered that the planets orbiting the young V\,1298 Tau exhibited significantly higher density than anticipated.
However, these detections have been challenged by the exoplanet community. \citet{Blunt2023} suggested that \citet{Suarez2022}'s detections may have been affected by over-fitting, thereby questioning the reliability of their RV Doppler detections.
This is crucial, especially in cases where the measurements are in tension with models.

\citet[][hereafter \citetalias{Bouma2020}]{Bouma2020} reported the discovery and validation of the young transiting exoplanet around \target. The star is a young ($\sim 35 {\rm Myr}$) F9/G0 dwarf star in the southern open cluster IC~2602. \target's main identifiers and parameters are given in Table~\ref{tab:parstellar}.
They used \tess\ cycle 1, the GAIA mission, multi-band photometry and spectroscopic observations to validate the planetary nature of a transit signal. The transit object corresponds to a planet with a radius of $\sim 0.77$\,\rjup\ and a period of $8.32$\,d. 
In this manuscript, we present a spectroscopic follow-up and exhaustive analysis of \target, to characterise further the nature of \targetb.
This paper is structured as follows: The photometric and spectroscopic data of  \target\ are detailed in Section~\ref{sec:data}. The analytical methods applied to this data are outlined in Section~\ref{sec:datanalaysis}. A discussion of the findings is provided in Section~\ref{sec:discusion}, and the paper concludes with a summary of the key outcomes in Section~\ref{sec:conclusions}. 
This manuscript is part of a series of papers under the project \emph{GPRV: Overcoming stellar activity in radial velocity planet searches} funded by the European Research Council (ERC, P.I.~S.~Aigrain). We acknowledge that while this paper was being reviewed, we learned that \citet{Damasso2024} were performing an independent analysis of the public HARPS data of \target. Should the reader wish to compare their results with ours, it would serve to further validate/test the findings presented in this manuscript.

\begin{table}
\caption{Main identifiers and parameters for \target.  \label{tab:parstellar} 
}
\begin{center}
\begin{tabular}{lcc} 
\hline
\noalign{\smallskip}
Parameter & Value &  Source \\
\noalign{\smallskip}
\hline
\noalign{\smallskip}
\multicolumn{3}{l}{\emph{Main identifiers}} \\
\noalign{\smallskip}
Gaia DR3  & 5251470948229949568 & \citet{Gaia2020}  \\
TYC & 8964-17-1 & \citet[][]{Hog2000} \\
2MASS & J10280898-6430189 &  \citet[][]{Cutri2003} \\
Spectral type & F9/G0 & \citet{Bouma2020} \\ 
\noalign{\smallskip}
\hline
\noalign{\smallskip}
\multicolumn{3}{l}{\emph{Equatorial coordinates, proper motion, and parallax}} \\
\noalign{\smallskip}
$\alpha$(J2000.0) &  10 28 08.9903 &  \citet{Gaia2020} \\
$\delta$(J2000.0) & -64 30 18.9364  & \citet{Gaia2020} \\
$\mu_\alpha$\,(mas\,${\rm yr^{{-1}}}$) & $ 	-17.912 \pm 0.014$ &  \citet{Gaia2020}  \\
$\mu_\delta$\,(mas\,${\rm yr^{{-1}}}$) & $ 11.490 \pm 0.014$ &   \citet{Gaia2020} \\
$\pi$\,(mas) & $ 	7.0108 \pm 0.0124$ & \ \citet{Gaia2020}  \\
Distance\,(pc) & $142.74 \pm 0.25$ & \citet{Gaia2020} \\
\noalign{\smallskip}
\hline
\noalign{\smallskip}
\multicolumn{3}{l}{\emph{Magnitudes}} \\
B &	$ 11.12 \pm 0.06$ &	 \citet[][]{Hog2000} \\  		
V &	$ 10.64 \pm 0.05$ &	  \citet[][]{Hog2000} \\ 
Gaia & $10.3598 \pm 0.0028 $ &	 \citet{Gaia2020}  \\ 	  		
J & $9.392 \pm 0.030$ &
	\citet{Cutri2003} 	 \\ 	  		
H &	$ 9.108 \pm 0.038$ &
	\citet{Cutri2003} 	 \\ 	  		
Ks &	$  8.933 \pm 0.026$ &
	\citet{Cutri2003} 	 \\
W1 & $8.901\pm0.023$ & AllWISE \\
W2 & $8.875\pm0.021$ & AllWISE \\
W3 & $8.875\pm0.020$ & AllWISE \\
\noalign{\smallskip}
\hline
\noalign{\smallskip}
\multicolumn{3}{l}{\emph{Stellar parameters}} \\
$T_{\rm eff}$ (K) & $5995\pm79$ & This work \\
\logg$_{\rm spec}$ (cgs) & $4.61\pm0.08$ & This work \\
{[Fe/H]} & $0.01\pm0.04$ & This work \\
\vsini\ (\kms) & $16.8\pm0.1$ & This work \\[1ex]
Mass (M$_\odot$) & $1.142^{+0.008}_{-0.011}$ & This work \\[1ex]
Radius (R$_\odot$) & $1.052^{+0.012}_{-0.007}$ & This work \\[1ex]
Luminosity (L$_\odot$) & $1.34^{+0.06}_{-0.12}$ & This work \\[1ex]
Density (\gcm) & $1.38^{+0.02}_{-0.04}$ & This work \\[1ex]
\logg$_{\rm isoc}$ (cgs) & $4.453^{+0.003}_{-0.008}$ & This work \\[1ex]
Age (Myr) & $35^{+7}_{-4}$ & This work \\[1ex]
$P_{\rm rot,max}$ (d) & $3.17\pm0.04$ & This work \\
\hline
\end{tabular}
\end{center}
\end{table}

\section{\target\ data}
\label{sec:data}

\subsection{\tess\ data}
\label{sec:tessdata}

\target\ (TIC 460205581) was observed by the \tess\ mission on cycles 1, 3 and 5. During cycle 1, \target\ was observed in Sectors 10 and 11 from 2019 March 26 to 2019 May 20 with a 2 min cadence. 
The \tess\ Science Processing Operations Center \citep[SPOC;][]{jenkins2016} transit search \citep{Jenkins2002,Jenkins2010,Jenkins2020} discovered a transiting signal with a period of 8.3~d in \target’s light curve.
This was announced in the \tess\ SPOC Data Validation Report \citep[DVR;][]{Twicken2018,Li2019} and designated by the TESS Science Office as TESS Object of Interest \citep[TOI;][]{Guerrero2021} \target.01 (hereafter \targetb). 

\oscar{
It should be noted that \target\ is situated in a relatively dense stellar region. Utilising Gaia DR2 data, \citetalias{Bouma2020} identified two sources brighter than $T=16$ within the same \tess\ pixel as \target: TIC 847769574 ($T=14.6$), which was argued a possible co-moving companion within the IC 2602 moving group, and TIC 460205587 ($T=13.1$), a background giant star.
\citetalias{Bouma2020} employed speckle imaging and photometric ground-base observations (refer to Section~\ref{sec:gbphotometry}) to rule out potential false-positive scenarios and confirm that the observed transits occur on \target, thereby validating the planetary nature of the transiting signal.
More recently, \citet{Behmard2022}, using Gaia DR3 data, suggested that TIC 847769574 is likely gravitational bound to \target. This companion is consistent with being an M dwarf, with an estimated mass of $0.393 \pm 0.034$,\msun\ and a projected separation of approximately 330\,AU, suggesting that the \target\ system may be binary, with TIC 847769574 being \target\,(B). However, the findings from \citetalias{Bouma2020} continue to support the scenario of the planet transiting \target\,(A). For simplicity, this manuscript will continue to refer to the planet as \targetb\ instead of \target\,(A)\,b.
We note that any extra flux coming from contaminating stars within \tess\ pixels (named crowding) is corrected for the PDCSAP light curves by the SPOC \citep{Stumpe2012}. Therefore, no further processing is needed for the analyses in this manuscript where we use PDCSAP light curves.
}

\begin{figure*}
    \centering
    \includegraphics[width=\textwidth]{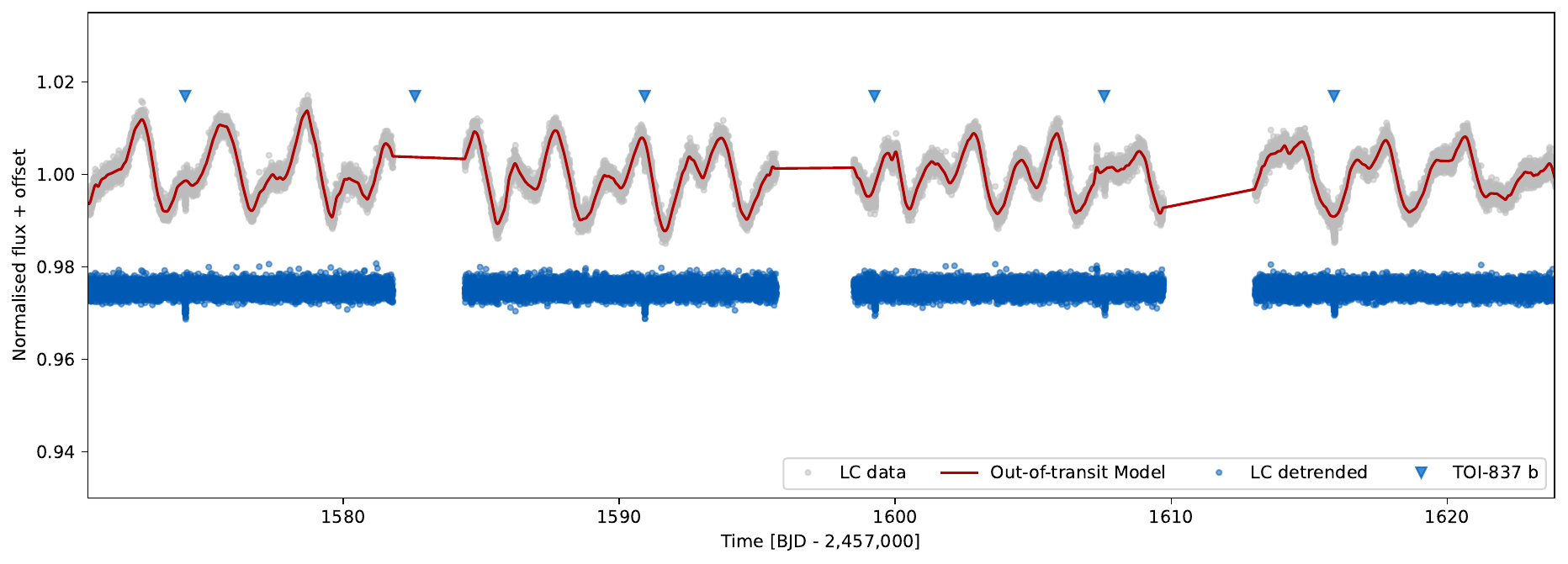} \\
    \includegraphics[width=\textwidth]{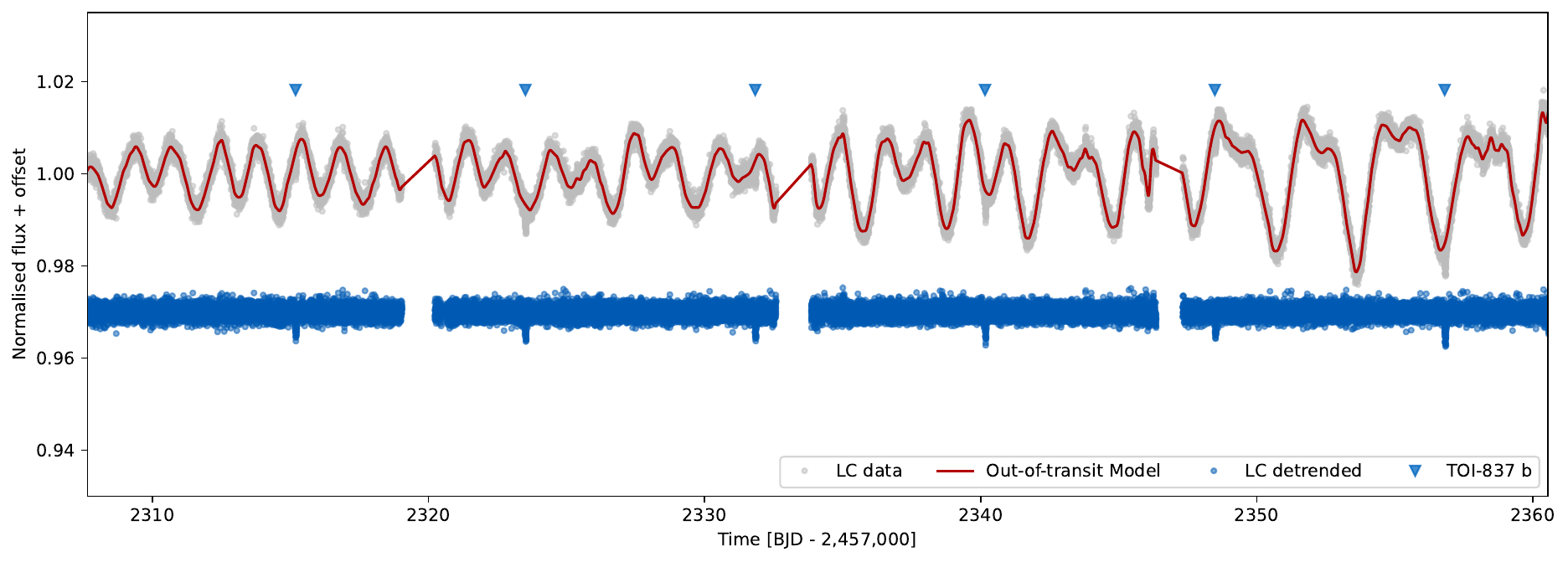} \\
    \includegraphics[width=\textwidth]{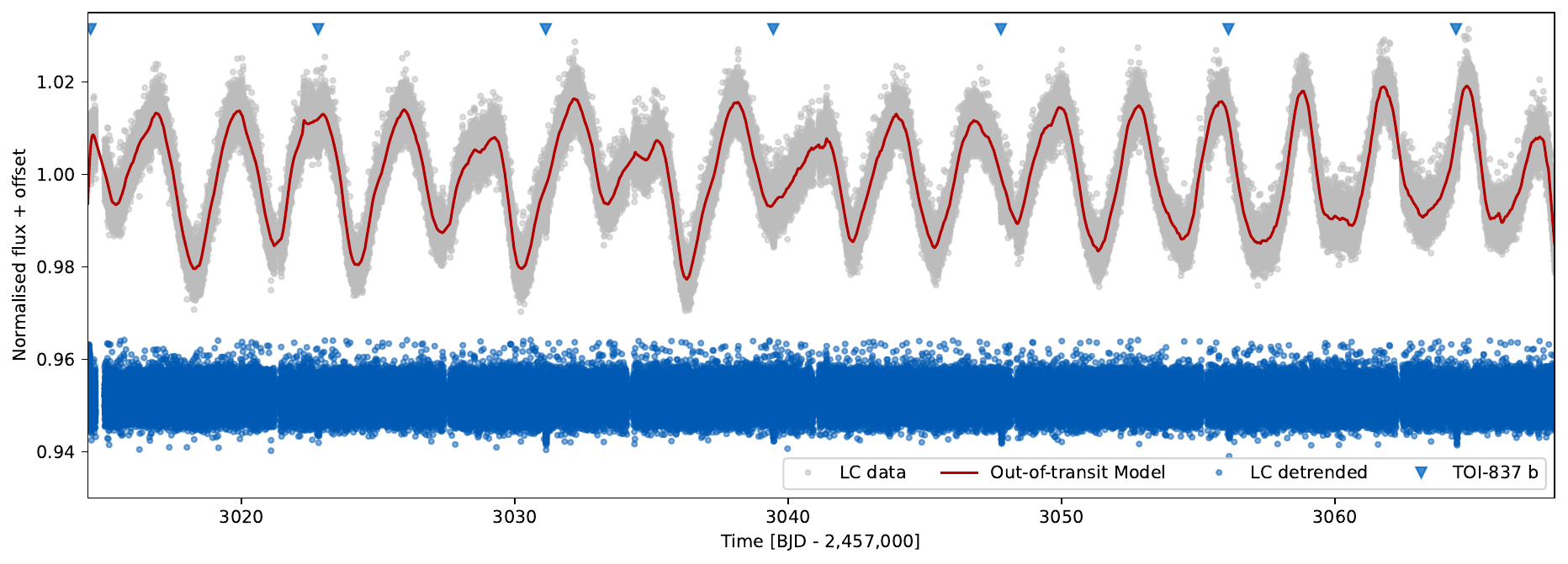} \\
    \caption{\tess\ light curves for \target\ in cycles 1, 3 and 5 (from top to bottom).
    \tess\ data are shown with grey points with the out-of-transit variability model over-plotted in red. 
    The resulting flattened light curve is presented with blue points.
    \target's transit positions are marked with a blue triangle.
    All plots have the same $y$-range to ease comparison of data and signals.
    }
    \label{fig:lc}
\end{figure*}

During cycles 3 and 5, \target\ was reobserved in Sectors 37 and 38 (from 2021 April 02 to 2021 May 26, with 2 min cadence) and Sectors 63 and 64 (2023 Mar 10 to 2023 May 04, with 20 s cadence), respectively. 
We downloaded the Presearch Data Conditioning Simple Aperture Photometry \citep[PDCSAP;][]{Smith2012,Stumpe2012,Stumpe2014} light curves for \target\ from the Mikulski Archive for Space Telescopes (MAST). Figure~\ref{fig:lc} shows the \target's light curves for all the three \tess\ cycles.
\target\ will be re-observed by \tess\ in Sector 90 during the first semester of 2025.

To perform the transit analysis we flattened the \tess\ light curves (see Sect.~\ref{sec:transitanalysis}). 
The \tess\ light curves for \target\ exhibit variations outside of transits, possibly due to stellar or instrumental variability. 
The \tess\ light curves were detrended using the open-source code \href{https://github.com/oscaribv/citlalicue}{\texttt{citlalicue} \faGithub} \citep{pyaneti2}. In essence, \citlalicue\ employs GPs, as implemented in \texttt{george} \citep[][]{george}, to model light curve fluctuations outside of transits. 
We provided \citlalicue\ with normalised light curves and the ephemeris for the transit signal. 
To focus on removing low-frequency signals, we used data binned at 3-hour intervals and masked all transits when fitting the GP with a Quasi-Periodic kernel \citep[as in][]{george}. 
We applied an iterative maximum Likelihood optimisation coupled with a $5$-sigma clipping technique to identify the best model for light curve variations outside of transits. Subsequently, we divided the entire light curves based on this model, resulting in a flattened curve showing only transit signals. It is worth noting that each \tess\ cycle was detrended individually. 
The detrended light curves for all \tess\ cycles are presented in Figure~\ref{fig:lc}.

\subsection{Ground-based photometric data}
\label{sec:gbphotometry}

\citetalias{Bouma2020} performed multi-band ground based follow-up of \target. For full details on the acquisition and reduction we refer the reader to that paper.
Four transit events were monitored using the 0.36-meter telescope at El Sauce Observatory, situated in Chile's Río Hurtado Valley. The data collection occurred during specific nights: the Cousins-R band was used on April 1 and 26 of 2020, the Cousins-I band on May 21, 2020, and the Johnson-B band on June 14, 2020. \target\ was also observed by the Antarctic Search for Transiting Exoplanets (ASTEP) telescope at the Concordia
base on the Antarctic Plateau, on the dates of May 12, May 29, June 14, and June 23 of 2020. We downloaded all these public data to include them in our analyses.

\subsection{HARPS spectroscopic observations}
\label{sec:harpsdata}

We acquired 75 high-resolution (R\,$\approx$\,115\,000, wavelength range of 380 to 690 nm) spectra of \target\ with the High Accuracy Radial velocity Planet Searcher \citep[HARPS;][]{Mayor2003} spectrograph. The instrument is mounted at the 3.6\,m ESO telescope at La Silla Observatory, Chile. 

The observations were carried out between October 03 2022 and January 16 2023, as part of the observing program 0110.C-4341(A) (PI: Yu). The typical exposure time per observation was 1800 s, except for the observations before October 25 where the exposure time was set to 900\,s. This produced spectra with a typical signal-to-noise (S/N) of 50 (30 for the 900s exposures) at 550 nm. 

We post-processed the 1-D spectra produced by the official data reduction software (DRS). All the spectra $S(\lambda_i,t)$ were first continuum normalised by \href{https://github.com/MichaelCretignier/Rassine_public}{\texttt{RASSINE} \faGithub} \citep{Cretignier2020b}, an upper envelope method fitting the spectra continua $C(\lambda_i,t)$. The normalised spectra time series $f(\lambda_i,t)=S(\lambda_i,t)/C(\lambda_i,t)$ was then post-processed by YARARA \citep{Cretignier2021}. Given the limited S/N of the observations, we restricted the cleaning by YARARA to cosmic rays and tellurics which are the only notable features at such a level of flux precision. 
From this step onwards, four observations were rejected by YARARA based on an anomalous number of detected outliers. Those spectra were all very low S/N observations ($<25$).
Once spectra were corrected of systematics, the normalised flux spectra $f(\lambda_i)$ were scaled back to absolute flux units spectra $S(\lambda_i)$ by using a reference continuum\footnote{Sometimes referred as a ``reference colour''.} $C_{\text{ref}}(\lambda)$ where the bolometric flux of the spectra was preserved during the scaling: $$S(\lambda_i,t) = f(\lambda_i,t) \cdot C_{\text{ref}}(\lambda_i)\cdot \frac{\sum_i C(\lambda_i)}{\sum_i C_{\text{ref}}(\lambda_i)}$$ This step is equivalent in practice to the colour correction usually performed on the order-by-order CCFs to remove any dependency with airmass or weather condition \citep{LovisPhD,CretignierPhD}.  We extracted the RVs and activity indicators by a Gaussian fit on CCFs obtained using the G2 mask of the HARPS DRS since for fast-rotating stars, the empirical line list \citep[see e.g.,][]{Cretignier2020} obtained from the observations could be imperfect due to a large number of blends. The full width at half maximum (FWHM) of the CCFs shows a consequent broadening with a value around FWHM $ \sim 26$ \kms\ in agreement with the fast rotation rate of the star (see Sect.~\ref{sec:stellarparameters}).
\oscar{
Since the CCFs are considerably broadened, we also extracted the RVs with a Generalised normal distribution (GND) as introduced in \citet[][]{Heitzmann2021}. GND has for advantage to contains an extra shape parameter $\beta$ that affects the kurtosis of the profile in order to produce a more "top-hat flatten" behaviour. When $\beta=2$, GND is fully equivalent to a Gaussian. We extracted the RVs using both a completely free GND model ($\beta(t)$ time dependent) and a more rigid one with the shape parameter $\beta$ fixed to the mean value ($\beta(t)=2.8$). The RVs obtained with these alternative parametrizations are fully consistent with the RVs obtained with the Gaussian fit. 
}
Our HARPS RV measurements have a typical error of 10\,\ms\ and a RMS of 100\,\ms. 
Table~\ref{tab:harps} lists the extracted HARPS RV, \oscar{RV GND, RV GND\_2.8}, FWHM, Bisector span (BIS),  calcium lines S-index (\sshk), and Hydrogen-alpha (\halpha) time series.
 
\begin{table*}
\begin{center}
\caption{HARPS spectroscopic measurements. The full version of this table is available in a machine-readable format as part of the supplementary material. \label{tab:harps}}
\begin{tabular}{cccccccccccccc}
\hline\hline
Time & RV & RV GND & RV GND\_2.8 & $\sigma_{\rm RV}$ & FWHM & $\sigma_{\rm FWHM}$ & BIS & $\sigma_{\rm BIS}$ & \sshk\ & $\sigma_{\rm S_{HK}}$ & \halpha\ & $\sigma_{\rm H_\alpha}$  \\
${\rm BJD_{TDB}}$ - 2\,470\,000 & \kms & \kms & \kms & \kms & \kms & \kms & \kms & \kms &  &  \\
\hline
2855.882062 & 0.0634 & 0.0497 & 0.0519 & 0.0160 & 24.0112 & 0.0521 & -0.0708 & 0.0216 & 0.0386 & 0.0242 & 0.3979 & 0.0039  \\ 
2859.873603 & -0.0075 & -0.0169 & -0.0167 & 0.0155 & 23.8725 & 0.0505 & -0.0957 & 0.0210 & 0.0811 & 0.0238 & 0.3918 & 0.0035  \\ 
2860.880897 & 0.0386 & -0.0272 & -0.0248 & 0.0206 & 23.7631 & 0.0671 & -0.4361 & 0.0278 & 0.0255 & 0.0347 & 0.3972 & 0.0048  \\ 
2861.882173 & 0.1588 & 0.1126 & 0.1105 & 0.0172 & 24.0269 & 0.0560 & -0.3255 & 0.0232 & 0.0683 & 0.0302 & 0.3954 & 0.0041  \\ 
2863.879982 & 0.0442 & -0.0012 & -0.0020 & 0.0152 & 23.9185 & 0.0493 & -0.3695 & 0.0204 & -0.0254 & 0.0214 & 0.3899 & 0.0031 \\ 

$\cdots$ \\
\hline
\end{tabular}
\end{center}
\end{table*}

\subsubsection{Periodograms}
\label{sec:periodograms}

\begin{figure}
    \centering
    \includegraphics[width=0.49\textwidth]{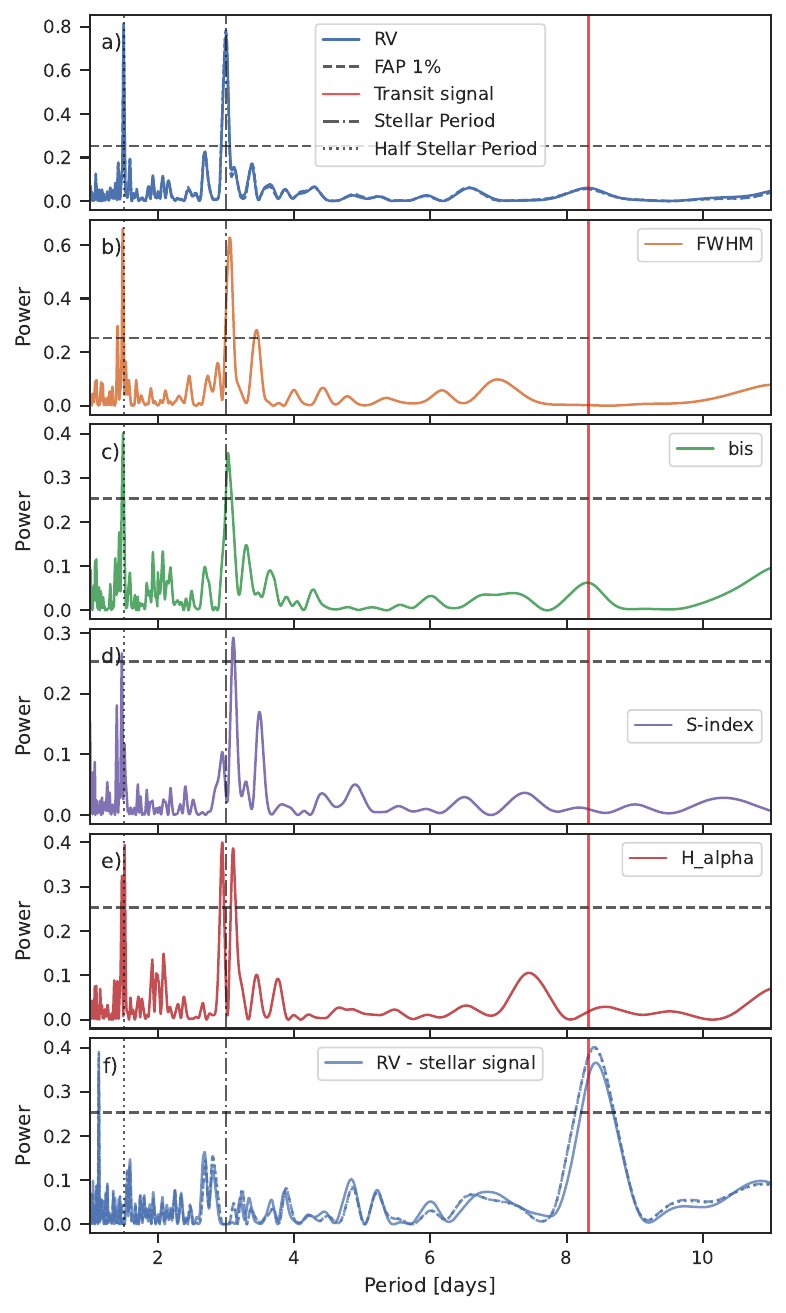}
    \caption{
    GLS periodograms of the spectroscopic time series. \oscar{GLS periodograms of the spectroscopic time series are presented in panels a) to f). Panel a) displays periodograms for Gaussian (solid line), GND (dotted line), and GND\_2.8 (dashed line) extracted RVs, which appear overlapped. Panel f) depicts a GLS periodogram for these RVs after removing the stellar model (see Sect~\ref{sec:rvanalysis}).} The horizontal dashed line indicates the 1\% False Alarm Probability (FAP). The dot-dashed and dotted black lines represent the stellar rotation period and its first harmonic, respectively, while the vertical red line marks the orbital period of \targetb.
    }
    \label{fig:periodiograms}
\end{figure}

As a first check to test the information contained in our HARPS spectroscopic time series, we ran a General Lomb-Scargle
\citep[GLS;][]{GLS} periodogram on them. 
Figure~\ref{fig:periodiograms} shows the periodogram of all the time series. We can see that all of the spectroscopic time series peak around $3$ and $1.5$\,d, which correspond to the rotational period of the star and its first harmonic.
The peaks at the stellar rotation period and its first harmonic suggest the presence of active regions on the stellar surface \citep[e.g.,][]{Aigrain2012}. No substantial peak corresponding to the orbital period of \targetb\ is evident in the raw RVs.
\oscar{
We also check for signals at longer periods (up to 100\,d, not shown in Fig.~\ref{fig:periodiograms}) and did not find any significant peak in any time series.
}

\section{Data analysis}
\label{sec:datanalaysis}

\subsection{Stellar parameters}
\label{sec:stellarparameters}

\subsubsection{Atmospheric parameters}

We used the HARPS spectra to derive the atmospheric parameters of \target. As this star is a fast rotator, we analysed the spectra using spectral synthesis. 
Two different methods were employed, \texttt{FASMA-synthesis}\footnote{\url{https://github.com/MariaTsantaki/FASMA-synthesis}} and \texttt{PAWS}\footnote{\url{https://github.com/alixviolet/PAWS}}, as well as two different sets of spectra, the DRS spectra and the YARARA-processed spectra (see Sect.~\ref{sec:harpsdata}).
Each set of spectra was shifted in the lab frame and co-added and the analyses were performed on this co-added spectrum. 

\texttt{FASMA-synthesis} made use of the MARCS atmospheric models \citep{MARCS} and the radiative transfer code \texttt{MOOG}\footnote{\url{https://www.as.utexas.edu/~chris/moog.html}}. Micro- and macroturbulent velocities were calculated through the calibration relations mentioned in \citet{Tsantaki-2013} and \citet{Doyle2014}, respectively. More details on this method can be found in \citet{Tsantaki2020}.

\texttt{PAWS} uses the functionalities from \texttt{iSPEC} \citep{Blanco-Cuaresma2019} and is described in more detail in \citet{Freckelton2024}. We used the Kurucz atmospheric models in our analysis \citep{Kurucz-93} and skipped the initial step of the equivalent width analysis as this is less reliable for fast rotators where the spectral lines are broadened. 

These four analyses provided values for the effective temperature, the surface gravity, the metallicity, and the projected rotational velocity. All values are in agreement with their 1-sigma errors. 
For our adopted parameters, listed in Table \ref{tab:parstellar}, we choose to take a weighted average of all the results. The high value of the projected rotational velocity (16.8 km/s) confirms the fast-rotating nature of \target. All other parameters are close to Solar values.

\subsubsection{Mass, radius, and age}

To calculate the fundamental stellar parameters, we made use of the \texttt{isochrones} \citep{Morton2015} package. Our stellar models came from the MESA Isochrones and Stellar Tracks \citep[MIST - ][]{Dotter2016} and we used the nested sampler \texttt{MultiNest} \citep{Feroz2019} for our likelihood analysis. As input data, we used the photometric magnitudes B, V, J, H, Ks, W1, W2, W3, the Gaia DR3 parallax, and the spectroscopically derived effective temperature and metallicity. In a similar way as described in \citet{Mortier2020}, we ran the code four times, each time changing the spectroscopic input to the individually obtained results and keeping the other data the same. Narrow bounds were set on the stellar age, between 30 and 46 Myr, following \citetalias{Bouma2020}, which helps constrain the stellar mass given strong correlations between mass and age for young stars.

After checking that all parameters from the individual runs agreed within errors, the final parameters and errors, listed in Table \ref{tab:parstellar}, were obtained from the median and 16th/84th percentile of the combined posterior distributions of the four runs. The mass and radius values are, unsurprisingly, close to Solar values. From these, we can recalculate a much more precise value of the surface gravity. It is lower than, but consistent with, the spectroscopically derived surface gravity. Using the stellar radius and the projected rotational velocity, we can furthermore also place an upper limit on the stellar rotation period. We find that the rotation period has a maximum value of $3.17\pm0.04$\,d. This is fully compatible with the values obtained with \tess\ and spectroscopic time series (see Sect.~\ref{sec:stellarsignal}).

\subsection{Transit analysis}
\label{sec:transitanalysis}

For all the subsequent planet analyses we used the code \pyaneti\ \citep{pyaneti,pyaneti2}. In all our runs we sample the parameter space with 250 walkers using the Markov chain Monte Carlo (MCMC) ensemble sampler algorithm implemented \oscar{ in \pyaneti\ \citep{pyaneti} which is based on \citet{emcee}}. 
The posterior distributions are created with the last 5000 iterations of converged chains. We thinned our chains by a factor of 10 giving a distribution of 125\,000  points for each sampled parameter. 

To speed up the transit modelling, we only model data chunks spanning a maximum of three hours on either side of each transit mid-time.
\tess\ and ground-base photometry were acquired with short cadence ($<2$\,min). Therefore we can assume that this data can be described by instantaneous evaluations of the transit models and we do not need model re-sampling \citep[see e.g.,][]{Gandolfi2018}. 
In total we have six different photometric data sets named: \tess\ 2 min data (cycles 1 and 3), \tess\ 25s data (cycle 5), ASTEP, and El Sauce R, I, and B bands.

To model the transits of \targetb, we need to set priors for the following set of parameters: time of mid-transit, $T_0$; orbital period, $P_{\rm orb}$; orbital eccentricity, $e$ and angle of periastron, $\omega$ using the polar parametrisation $\sqrt{e} \cos \omega_\star$ and $\sqrt{e} \sin \omega_\star$; scaled planetary radius $R_{\rm p}/R_\star$; the stellar density, $\rho_\star$; and the  the limb darkening parameters $q_1$ and $q_2$ for each band \citep[following][models and parametrisations]{Mandel2002,Kipping2013}. 
We assume a black object transit and we sample for a single $R_{\rm p}/R_\star$ for all bands (\citetalias{Bouma2020} showed that there is no variation of $R_{\rm p}/R_\star$ in different bands). 
The scaled semi-major axis, $a/R_\star$, is recovered from $\rho_\star$ and Kepler's third law \citep[see e.g.,][]{Winn2010}. 
The model also includes photometric jitter term per data set to penalise the likelihood.
We start by assuming that the orbit is circular, so we fix $\sqrt{e} \cos \omega_\star = \sqrt{e} \sin \omega_\star = 0$. For the rest of parameters we set wide informative uniform priors as listed in Table~\ref{tab:pars}.

To start our transit analysis, it is worth mentioning that \targetb's transit looks V-shaped, indicating that the transit could be grazing (where the grazing condition is given when $ b > 1 - R_{\rm p}/R_\star$).
\citetalias{Bouma2020} reported $b = 0.94 \pm 0.013$ and $R_{\rm p}/R_\star = 0.08 \pm 0.01$, suggesting that the transit is either grazing or nearly grazing.
We note that \citetalias{Bouma2020} performed this analysis using the available data at the time: TESS cycle 1 and the ground-based data. 
We performed an analysis using exactly this data set, and we obtained fully consistent results of $b = 0.923_{-0.011}^{+0.019}$ and $R_{\rm p}/R_\star = 0.079_{-0.003}^{+0.006}$. The small differences with \citetalias{Bouma2020} can be explained by the different ways to deal with the detrending and/or different parametrisations.

\begin{figure}
    \centering
    \includegraphics[width=0.45\textwidth]{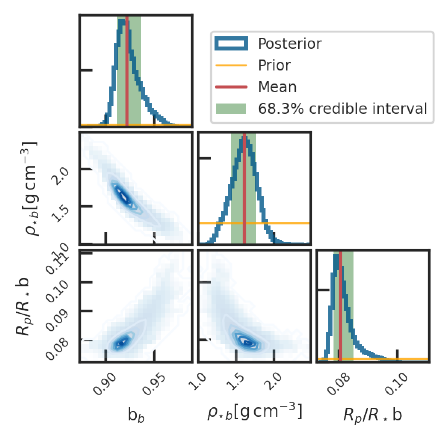}
    \caption{Correlation plot for the main parameters of the transit analysis. }
    \label{fig:trcorr}
\end{figure}

We then repeated the analysis including the new data of the \tess\ cycles 3 and 5. We obtained more precise values of $b = 0.921_{-0.009}^{+0.014}$ and $R_{\rm p}/R_\star = 0.080 _{-0.002}^{+0.004}$. 
These values are still consistent with a grazing transit. 
Figure~\ref{fig:trcorr} shows a correlation plot between $b$, $\rho_\star$, and $R_{\rm p}/R_\star$. We can see that the grazing condition covers only the low probability tail of the posterior distribution.
Therefore, we are able to put strong constraints on the inferred planetary radius.

\begin{figure}
    \centering
    \includegraphics[width=0.45\textwidth]{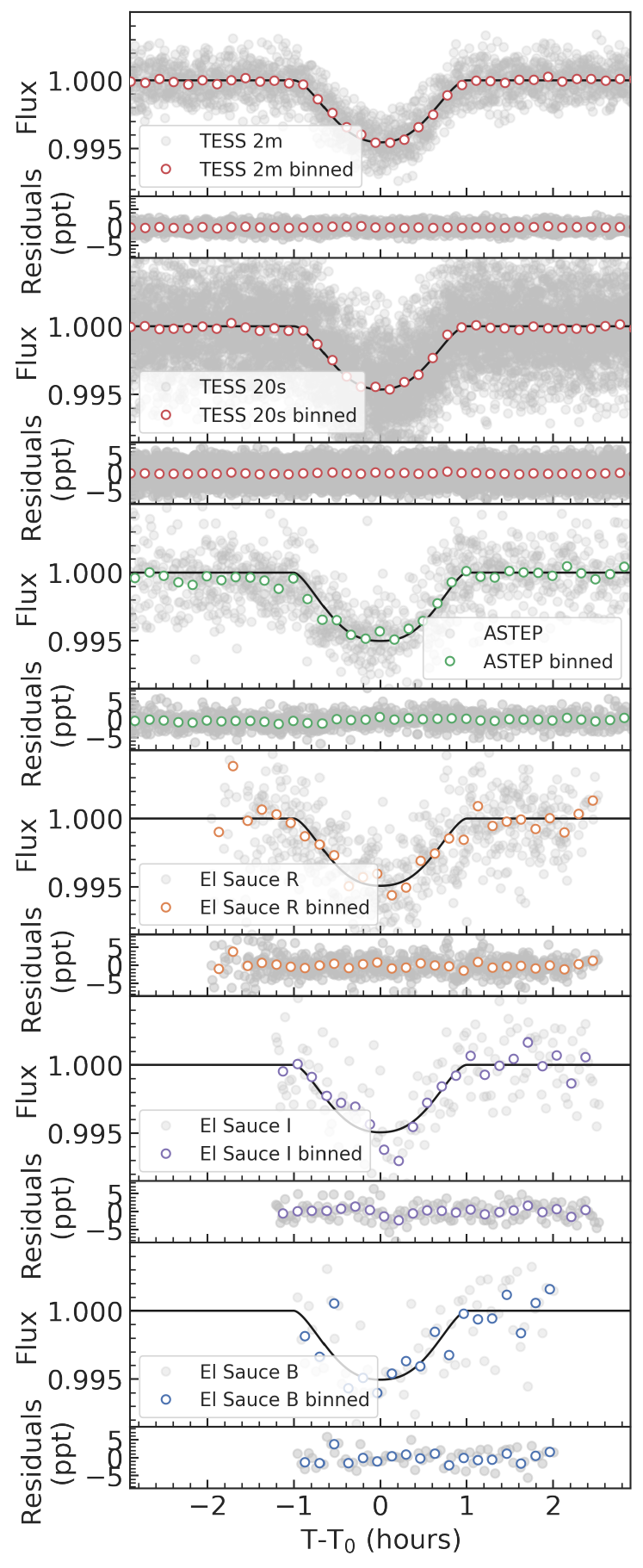}
    \caption{Phase-folded light curves of \targetb\ for different data sets. Nominal observations are shown in light grey. Coloured circles represent 10-min binned data. Transit models are shown with a solid black line.}
    \label{fig:transits}
\end{figure}

It is also worth mentioning that the stellar density recovered from the transit analysis, $\rho_\star = 1.58 \pm 0.17$\,\gcm, is fully consistent with the stellar parameters derived in Section~\ref{sec:stellarparameters}. This suggests that the planetary orbit is nearly circular.
To test this further, we performed another transit analysis where we used a Gaussian prior on the stellar density using the value in Table~\ref{tab:parstellar}. We also sample for the orbit eccentricity using the polar parametrisation $\sqrt{e} \cos \omega_\star$ and $\sqrt{e} \sin \omega_\star$. 
We inferred $\sqrt{e} \cos \omega_\star = 0.08_{-0.15}^{+0.12}$ and $\sqrt{e} \sin \omega_\star = 0.10_{-0.35}^{+0.39}$ that relates to an eccentricity of $e = 0.10_{-0.06}^{+0.17}$. These results are consistent with a nearly circular orbit for \targetb.
It is also worth mentioning that the model with a circular orbit is strongly preferred over the model with eccentricity with a difference of Akaike Information Criterion ($\Delta$\,AIC) of 52.
We then conclude that the photometric data strongly prefers a model assuming a circular orbit.
Figure~\ref{fig:transits} shows \targetb's transits for the different data sets with the respective inferred models.

\begin{table*}
\begin{center}
  \caption{Modelled and derived parameters for \targetb. \label{tab:pars}} 
  \begin{tabular}{lcc}
  \hline
  \hline
  \noalign{\smallskip}
  Parameter & Prior$^{(a)}$ & Final value$^{(b)}$ \\
  \noalign{\smallskip}
  \hline
  \noalign{\smallskip}
  \multicolumn{3}{l}{\emph{\bf \targetb's parameters }} \\
  \noalign{\smallskip}
    Orbital period $P_{\mathrm{orb}}$ (days)  & $\mathcal{U}[8.3239 , 8.3259  ]$ &\Pb[] \\
    Transit epoch $T_0$ (BJD$_\mathrm{TDB}-$2\,457\,000)  & $\mathcal{U}[2356.77 , 2356.87]$ & \Tzerob[]  \\  
    Scaled planet radius $R_\mathrm{p}/R_{\star}$  &$\mathcal{U}[0.0,0.2]$ & \rrb[]  \\
    Impact parameter, $b$ & $\mathcal{U}[0,1.2]$ & \bb[] \\
    $\sqrt{e} \cos \omega_\star$ & $\mathcal{F}[0]$ & 0 \\
    $\sqrt{e} \sin \omega_\star$ & $\mathcal{F}[0]$ & 0 \\
    Doppler semi-amplitude, $K$ (\ms) & $\mathcal{U}[0, 100]$ & \kb[] \\
    \multicolumn{3}{l}{\emph{ \bf GP hyperparameters}} \\
   GP Period $P_{\rm GP}$ (days) &  $\mathcal{U}[2.5,3.5]$ & \jPGP[] \\
    $\lambda_{\rm p}$ &  $\mathcal{U}[0.1,5]$ &  \jlambdap[] \\
    $\lambda_{\rm e}$ (days) &  $\mathcal{U}[1,100]$ &  \jlambdae[] \\
    $A_{\rm RV}$ (\ms)  &  $\mathcal{U}[0,1000]$ & \AGP[] \\
    \multicolumn{3}{l}{\emph{ \bf Other parameters}} \\
    Stellar density $\rho_\star$ (\gcm) & $\mathcal{U}[0.5,2.5]$ & \denstrb[] \\ 
    \tess\ Parameterised limb-darkening coefficient $q_1$  &$\mathcal{U}[0,1]$ & \qoneS[] \\ 
    \tess\ Parameterised limb-darkening coefficient $q_2$  &$\mathcal{U}[0,1]$ & \qtwoS[] \\ 
    El Sauce-R Parameterised limb-darkening coefficient $q_1$  &$\mathcal{U}[0,1]$ & \qoneelsaucer[] \\ 
    El Sauce-R Parameterised limb-darkening coefficient $q_2$  &$\mathcal{U}[0,1]$ & \qtwoelsaucer[] \\ 
    El Sauce-I Parameterised limb-darkening coefficient $q_1$  &$\mathcal{U}[0,1]$ & \qoneelsaucei[] \\ 
    El Sauce-I Parameterised limb-darkening coefficient $q_2$  &$\mathcal{U}[0,1]$ & \qtwoelsaucei[] \\ 
    El Sauce-B Parameterised limb-darkening coefficient $q_1$  &$\mathcal{U}[0,1]$ & \qoneelsauceb[] \\ 
    El Sauce-B Parameterised limb-darkening coefficient $q_2$  &$\mathcal{U}[0,1]$ & \qtwoelsauceb[] \\ 
    ASTEP Parameterised limb-darkening coefficient $q_1$  &$\mathcal{U}[0,1]$ & \qoneastep[] \\ 
    ASTEP Parameterised limb-darkening coefficient $q_2$  &$\mathcal{U}[0,1]$ & \qtwoastep[] \\ 
    Jitter term $\sigma_{\rm TESS, 2 min}$ (ppm) & $\mathcal{J}[1,5000]$ & \jtrtessm[] \\
    Jitter term $\sigma_{\rm TESS, 25 s}$ (ppm) & $\mathcal{J}[1,5000]$ & \jtrtesss[] \\
    Jitter term $\sigma_{\rm R}$ (ppm) & $\mathcal{J}[1,5000]$ & \jtrelsaucer[] \\
    Jitter term $\sigma_{\rm I}$ (ppm) & $\mathcal{J}[1,5000]$ & \jtrelsaucei[] \\
    Jitter term $\sigma_{\rm B}$ (ppm) & $\mathcal{J}[1,5000]$ & \jtrelsauceb[] \\
    Jitter term $\sigma_{\rm ASTEP}$ (ppm) & $\mathcal{J}[1,5000]$ & \jtrastep[] \\
    HARPS offset (\kms) & $\mathcal{U}[-0.67,0.71]$ & \HARPS[] \\
    HARPS Jitter, $\sigma_{\rm HARPS}$ (\ms)  & $\mathcal{J}[1,100]$ & \jHARPS[] \\
    \noalign{\smallskip}
    \hline
    \multicolumn{3}{l}{\emph{\bf \targetb's derived parameters }} \\
    Planet radius, ($R_{\rm J}$)  & $\cdots$ & \rpb[] \\
    Planet mass, ($M_{\rm J}$)  & $\cdots$ & \mpb[] \\
    Planet density, $\rho_{\rm p}$ (g\,cm$^{-3}$) & $\cdots$ &  \denpb[] \\
    Scaled semi-major axis $a/R_\star$ & $\cdots$ & \arb[] \\
    semi-major axis $a$ (AU)  & $\cdots$ & \ab[] \\
    Orbital inclination $i$ (deg)  & $\cdots$ & \ib[] \\
     Equilibrium temperature$^{(\mathrm{c})}$ $T_{\rm eq}$ ($K$)   & $\cdots$ & \Teqb[]  \\
    Transit duration $\tau_{\rm d}$ (h) & $\cdots$ & \ttotb[] \\
    Planet surface gravity $g_{\rm p}$ (${\rm cm\,s^{-2}}$)$^{(d)}$   & $\cdots$ & \grapb[]  \\
    Planet surface gravity $g_{\rm p}$ (${\rm cm\,s^{-2}}$)$^{(e)}$   & $\cdots$ & \grapparsb[] \\
    Insolation $F_{\rm p}$ ($F_{\oplus}$)   & $\cdots$ & \insolationb[] \\
    \hline
\multicolumn{3}{l}{\footnotesize $^a$ $\mathcal{F}[a]$ refers to a fixed value $a$, $\mathcal{U}[a,b]$ to an uniform prior between $a$ and $b$, $\mathcal{N}[a,b]$ to a Gaussian prior with mean $a$ and standard deviation $b$,}\\
\multicolumn{3}{l}{ and $\mathcal{J}[a,b]$ to the modified Jeffrey's prior as defined by \citet[eq.~16]{Gregory2005}.}\\
\multicolumn{3}{l}{\footnotesize $^b$ Inferred parameters and errors are defined as the median and 68.3\% credible interval of the posterior distribution.}\\
\multicolumn{3}{l}{\footnotesize $^c$ Assuming an albedo of zero.} \\
\multicolumn{3}{l}{$^d$ Derived using $g_{\rm p} = G M_{\rm p} R_{\rm p}^{-2}$. } \\
\multicolumn{3}{l}{$^e$ Derived using sampled parameters following \citet{Sotuhworth2007}.}\\
  \end{tabular}
\end{center}
\end{table*}

\subsection{Characterising the stellar signal}
\label{sec:stellarsignal}

Because of its youth, \target\ has strong stellar signals in its light curve and RVs time series. 
We performed several analyses of the light curves and spectroscopic time series to analyse the time scales over which the stellar signal evolves. 

\subsubsection{Stellar signal in the \tess\ light curves}
\label{sec:lcgps}

We first performed a onedimensional GP regression of the \tess\ light curves to study the behaviour of the stellar signal at the different \tess\ cycles.
Since we are interested in modelling the long-term evolution of the light curve, we binned the data into 3 hour chunks and masked the transits out of the light curve. 

We model the covariance between two data points at times $t_i$ and $t_j$ for each time-series as
\begin{equation}
    \gamma_{\rm 1D} = A^2 \gamma_{i,j},
\end{equation}
\noindent
where $A$ is an amplitude term, and $\gamma_{i,j}$ is the Quasi-Periodic (QP) kernel given by
\begin{equation}
   \gamma_{{\rm QP},i,j} = \exp 
    \left[
    - \frac{\sin^2[\pi(t_i - t_j)/P_{\rm GP}]}{2 \lambda_{\rm P}^2}
    - \frac{(t_i - t_j)^2}{2\lambda_{\rm e}^2}
    \right],
    \label{eq:gamma}
\end{equation}
\noindent
whose hyperparameters are: \pgp, the GP characteristic period; \lbp, the inverse of the harmonic complexity; and \lbe, the long-term evolution timescale. 

We sample 5 parameters in each run: four GP hyperparameters ($A$, \pgp, \lbe, \lbp), and a jitter term included in the Gaussian Likelihood. 
We set wide uniform priors for all the parameters: for \lbe\ we set a uniform prior between 0.1 and 100~d, for \lbp\ between 0.1 and 5, and for \pgp\ between 2.5 and 3.5~d.
We sample the parameter space using the built-in MCMC sampler in \pyaneti\ using the same criteria as in Sect.~\ref{sec:transitanalysis}. 

\begin{figure}
    \centering
    \includegraphics[width=0.45\textwidth]{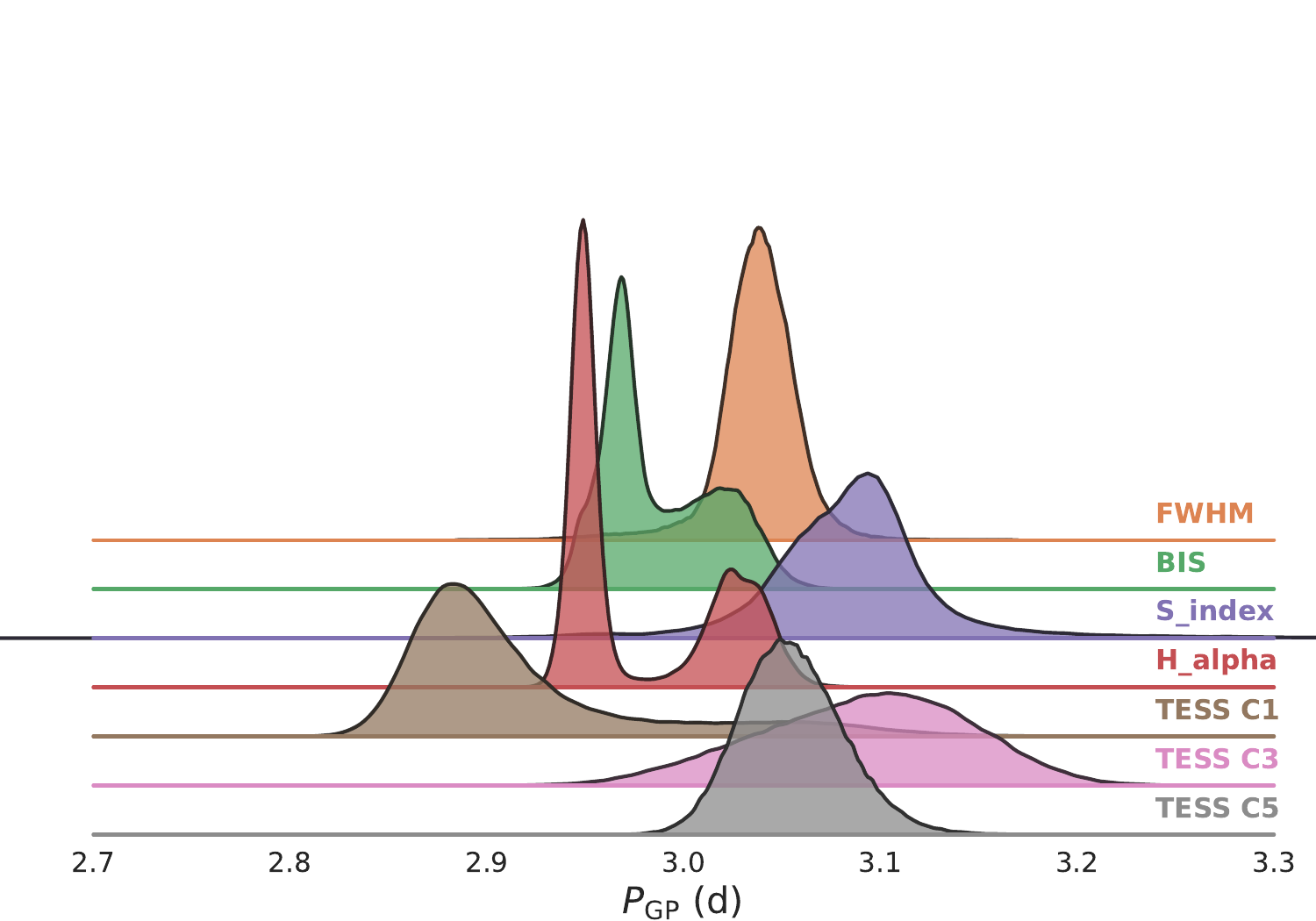}\\
    \includegraphics[width=0.45\textwidth]{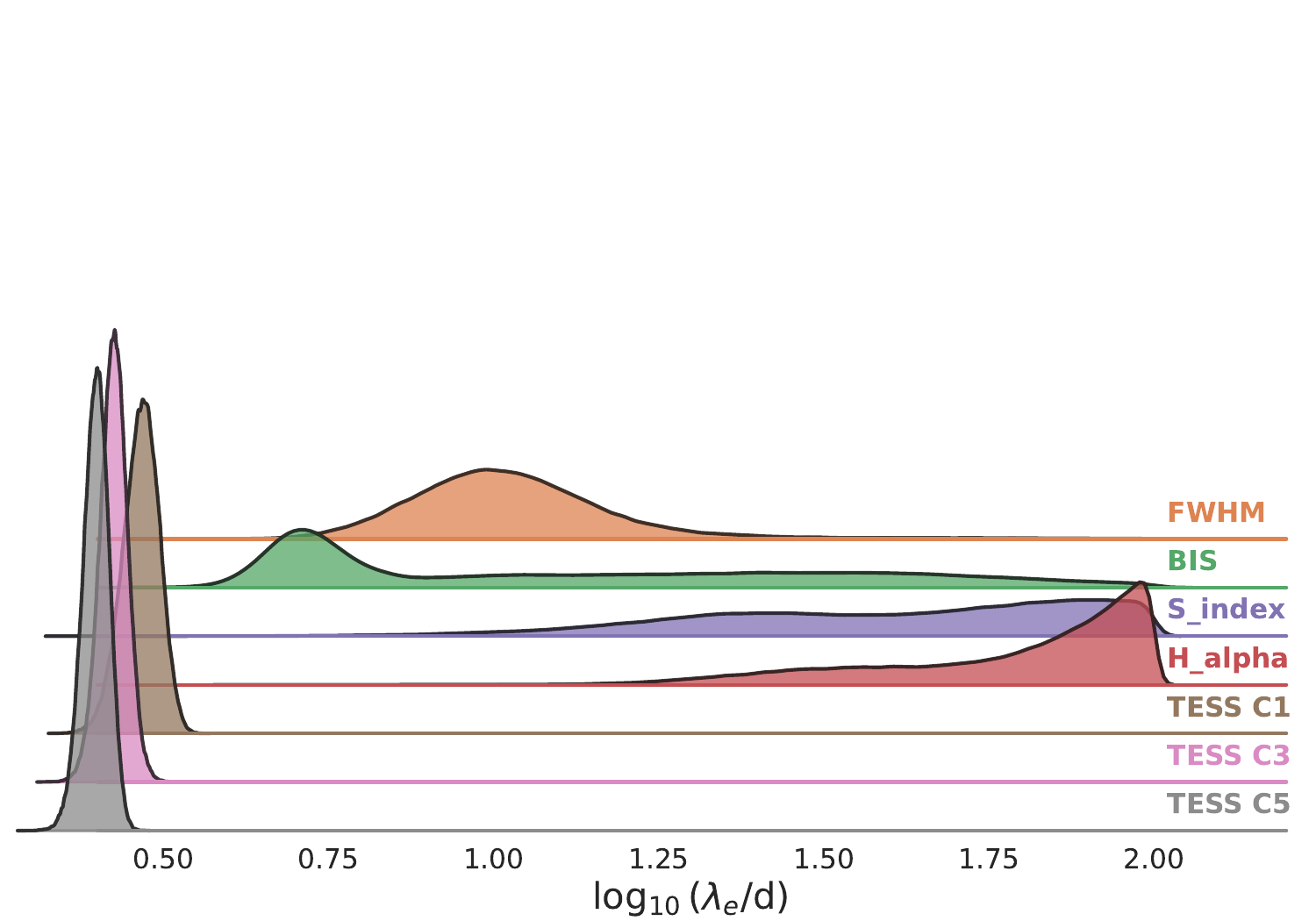}\\
    \includegraphics[width=0.45\textwidth]{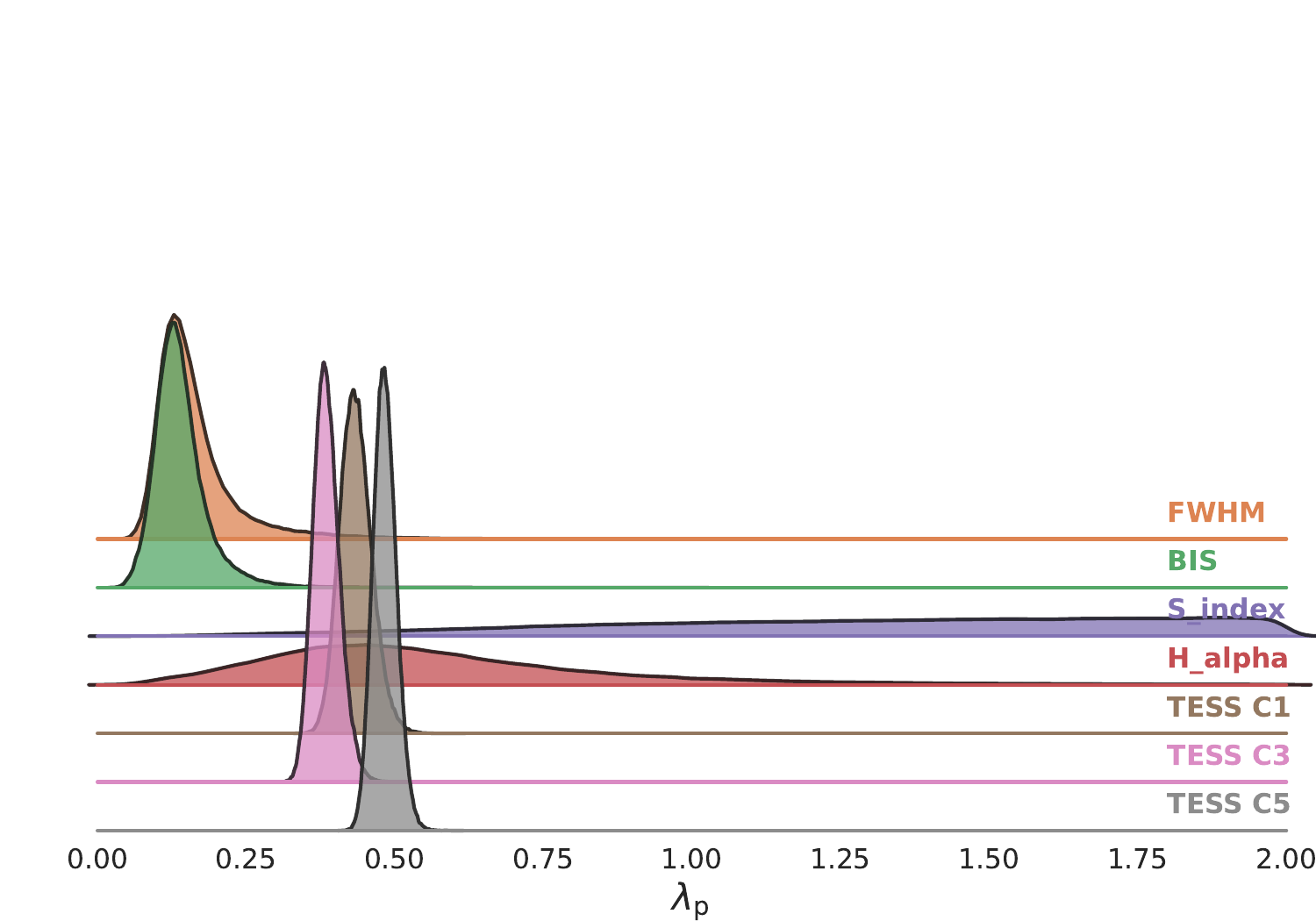}
    \caption{Posterior distributions for \pgp (top), $\log_{10}($\lbe$/{\rm d})$ (middle), and \lbp\ (bottom). Results for FWHM (orange), BIS (green), \sshk\ (purple), \halpha\ (red), and \tess\ cycles 1 (brown), 3 (pink), 5 (grey) are shown for each sub-panel.}
    \label{fig:posteriors1gp}
\end{figure}

\begin{figure*}
    \centering
    \includegraphics[width=\textwidth]{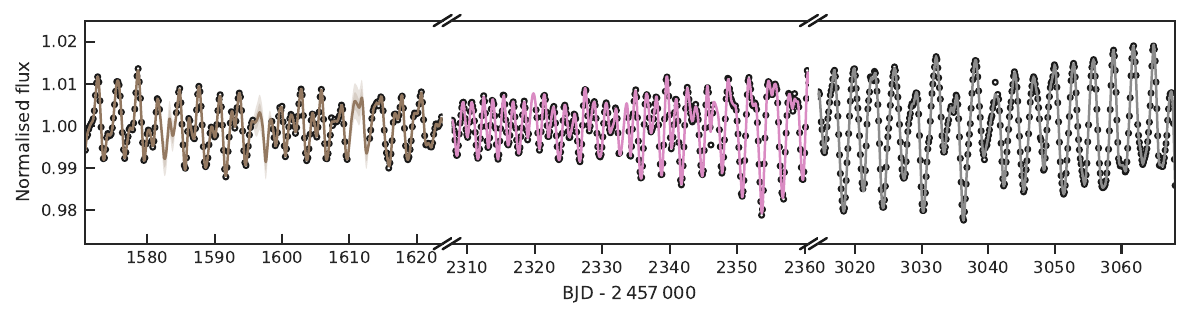}\\
    \includegraphics[width=\textwidth]{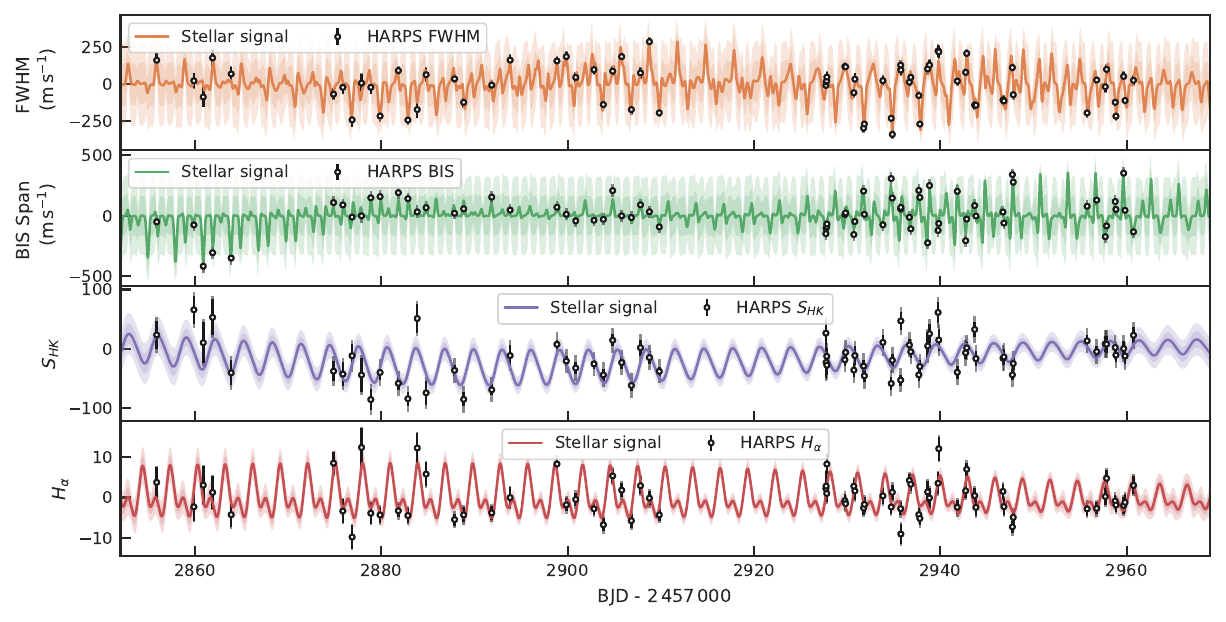}
    \caption{
    \emph{Top panel:} \target's TESS light curves and GP inferred models. 
    The corresponding measurements (3\,h binned data) are shown with black circles. 
    Solid-coloured lines show the corresponding inferred signal coming from our GP regression for \tess\ Cycles 1 (brown), 3 (pink), and 5 (grey).
    \emph{Bottom panel:} \target\ spectroscopic time-series for FWHM, BIS Span, \sshk, and \halpha. 
    The corresponding measurements are shown with black circles with error bars with a semi-transparent error bar extension accounting for the inferred jitter. 
    Solid-coloured lines show the corresponding inferred signal coming from our GP regression, while light-coloured shaded areas show the one and two-sigma credible intervals of the corresponding GP model.
    }
    \label{fig:timeseries1dgp}
\end{figure*}

Table~\ref{tab:1gphp} shows the inferred \pgp, \lbe, and \lbp\ hyperparameters for all the light curves. Figure~\ref{fig:posteriors1gp} shows the inferred posterior distributions for the \pgp, \lbe, and \lbp\ parameters for all the modelled time series.
Figure~\ref{fig:timeseries1dgp} shows the \tess\ light curves with the inferred GP model.
The first thing we observe is that the recovered \pgp\ is consistent within the maximum rotational period derived in Section~\ref{sec:stellarparameters}.
However, the rotational period obtained from \tess\ Cycle 1 is consistent with Cycles 3 and 5 just at 3 sigma. This could be evidence of differential rotation. This can also be caused by the relatively short evolution time-scale recovered that is smaller than the inferred period \citep[see e.g.,][]{Rajpaul2015}.  
We note that all the hyperparameters seem consistent between the different light curves. This is unusual for young stars given that each TESS cycle happens two years after another and the evolution of stellar activity can manifest as different signals that can be explained by different GP hyperparameters \citep[e.g.,][]{Barragan2021b}.

\begin{table}
\begin{center}
\caption{Recovered hyperparameters for onedimensional GP regression for different stellar time series. \label{tab:1gphp}} 
\begin{tabular}{lccccc}
\hline\hline
Time-series  & \pgp\ [d] & \lbe\ [d] & \lbp\   \\
\hline
\tess\ cycle 1 &  $2.90 _{ - 0.03 } ^ { + 0.07 }$ & $2.94 _{ - 0.19 } ^ { + 0.18 }$ & $0.43 _{ - 0.03 } ^ { + 0.03 }$\\
\tess\ cycle 3 &  $3.09 _{ - 0.06 } ^ { + 0.05 }$ & $2.66 _{ - 0.12 } ^ { + 0.13 }$ & $0.38 _{ - 0.02 } ^ { + 0.02 }$\\
\tess\ cycle 5 &  $3.05 _{ - 0.02 } ^ { + 0.03 }$ & $2.50 _{ - 0.12 } ^ { + 0.11 }$ & $0.48 _{ - 0.02 } ^ { + 0.02 }$\\
\hline
FWHM & $3.04 \pm 0.02$ & $10.3_{-2.6}^{+4.1}$ & $0.15_{-0.04}^{+0.06}$ \\
BIS Span & $2.98_{-0.02}^{+0.05} $ & $11.7_{-6.7}^{+30.4}$  & $0.13_{-0.03}^{+0.04}$  \\
\sshk & $3.09 \pm 0.03$ & $44.3_{-24.9}^{+35.4}$ & $1.35_{-0.55}^{+0.45}$ \\
\halpha & $2.95_{-0.01}^{+0.08}$ & $71.8_{-36.5}^{+20.8}$ & $0.51_{-0.21}^{+0.31}$ \\
\hline
\end{tabular}
\end{center}
\end{table}

\subsubsection{Stellar signal in the activity indicators}
\label{sec:ssactivityindicators}

We performed an individual analysis of each activity indicator as the one presented in \citet{Barragan2023}. We analyse the FWHM, BIS, \sshk, and \halpha\ spectroscopic time series. 
Note that, in comparison with Section~\ref{sec:lcgps}, we are now facing a more difficult signal characterisation given that the spectroscopic time series are not as well sampled as the \tess\ light curves.

In Section~\ref{sec:periodograms} we show how all the spectroscopic time-series periodograms show significant peaks around 3 and 1.5\,d, which are related to the stellar rotational period and its first harmonic, respectively.
To better characterise the periodic signals and analyse further signal complexity, we performed a GP regression on each time series using the 1D modelling setup as the one presented in Sect.~\ref{sec:lcgps}.
We sample for 6 parameters in each run, the kernel hyperparameters ($A$, \pgp, \lbe, \lbp), an offset, and a jitter term. 
The priors used were uniform with the ranges: for \lbe\ between 0.1 and 100~d (that corresponds to the HARPS observing window), \lbp\ between 0.1 and 2, and for \pgp\ between 2.5 and 3.5\,d.

Table~\ref{tab:1gphp} and Figure~\ref{fig:posteriors1gp} summarise the inferred GP hyperparameters. 
Figure~\ref{fig:timeseries1dgp} shows the spectroscopic time-series data together with the inferred GP model.
The first thing we note is that all activity indicators can recover a period that is consistent with the stellar rotation period. This is consistent with the results found in the periodogram analysis (even the double peak observed in the \halpha\ periodogram is presented in the recovered posterior, see Sect.~\ref{sec:periodograms}). 
We were expecting a better agreement between the hyperparameters recovered from the spectroscopic time series \citep[see][]{Barragan2023}. 
From Figure~\ref{fig:timeseries1dgp} we can see by eye that the recovered process in each time series looks significantly different. The processes recovered to explain FWHM and BIS Span present variations that are shorter than the time sampling of the HARPS data, this behaviour can be a result of overfitting \citep[see][]{Blunt2023}.
We can also see that the recovered signals for \sshk\ and \halpha\ are consistent with a low harmonic complexity process. 
\citet{Barragan2023} noted that this behaviour can be caused by relatively high white noise that does not allow to characterise the complexity of the underlying signal.

\oscar{
These results suggest that the data cadence is not good enough to  characterise the stellar signal in the activity indicators.}
The rotation period of the star of 3\,days is short compared with our sampling strategy of taking one point per night. Three points may not be enough to sample the complexity of the stellar signal, especially if this signal is expected to vary significantly in short time scales, as suggested by the \tess\ photometry. Therefore, our GP regression may not be able to characterise in full the stellar signal.  
Furthermore, the rotation period of the star is close to the integer of 3\,d. This has as a consequence that observations do not sample properly the phase of the stellar rotation given that observations happen at similar times during the whole observing run. 
This is worsened in the case in which the stellar signal evolves drastically between each period.

We conclude that for our spectroscopic time series, we can constrain the rotation period of the star. However, any further complexity pattern that may exist cannot be characterised by our dataset.
\oscar{
Because of this lack of information on the activity indicators, we conclude that they do not bring enough information of the stellar signal into a multidimensional GP regression analysis. In Appendix~\ref{sec:appendix} we show the results and a discussion on the results of applying a multidimensional GP framework to this data set.
}

\subsection{RV analysis}
\label{sec:rvanalysis}

We also performed onedimensional GP modelling to the RVs time series. 
We first ran a model identical to the GP analyses in Sect.~\ref{sec:ssactivityindicators}. This means that we sample for the QP kernel hyperparameters ($A$, \pgp, \lbe, \lbp), and a jitter term.
As a mean function, we included just an offset (with no Keplerian model yet). For the GP hyperparameters, we use the same priors as the ones described in Sect.~\ref{sec:ssactivityindicators}.
\oscar{
We run this analysis for the Gaussian, GND and GND\_2.8 RV time series.
For the Gaussian RVs we recover the hyperparameters \pgp\,$=3.00 \pm 0.02$\,d, \lbe\,$=42_{-16}^{+25}$\,d, and \lbp\,$=0.39_{-0.13}^{+0.22}$; for GND \pgp\,$=3.00 \pm 0.01$\,d, \lbe\,$=69_{-24}^{+22}$\,d, and \lbp\,$=0.42_{-0.14}^{+0.26}$; and GND\_2.8 \pgp\,$=3.00 \pm 0.01$\,d, \lbe\,$=67 \pm 23$\,d, and \lbp\,$=0.42_{-0.14}^{+0.25}$.
We can see that the hyperparameters are consistent between the different RV flavours.}
We can also see that besides the period, the hyperparameters do not compare with any of the hyperparameters obtained from the other spectroscopic time series (see Table~\ref{tab:1gphp}).

We then repeated the analysis but this time we added a mean function that includes an offset and a Keplerian model to account for the Doppler effect of \targetb. For the planet model we assume a circular orbit, we set Gaussian priors for the planet ephemeris based on the analysis in Sect.~\ref{sec:transitanalysis}. For the Doppler semi-amplitude we set a uniform prior between 0 and 100\,\ms. 
\oscar{
We inferred hyperparameters fully consistent with the no-planet models. 
We also recovered a Doppler semi-amplitude of $K = 36.6 \pm 6.5$\,\ms for the Gaussian RVs, $K = 34.7 \pm 5.6$\,\ms\ for the GND RVs, and $K = 34.6 \pm 5.5$\,\ms\ for the GND\_2.8 time series. All these values are fully consistent between them.
}

\oscar{
When comparing all models, the ones with the planetary signal are strongly favoured with a $\Delta\,{\rm AIC} > 20$. 
It is also worth mentioning that the recovered jitter term for the model with the planet signal is significantly smaller than the one without a Keplerian signal. The jitter values reduce from 38\,\ms\ to 30\,\ms, from 37\,\ms\ to 27\,\ms, and 36\,\ms\ to 26\,\ms, for the Gaussian, GND and GND\_2.8 time series, respectively. 
}
This is also evidence that the model that better adjusts the data is the one where the planetary signal is included.
\oscar{
The GND\_2.8 RV time series yields the most precise Doppler semi-amplitude and exhibits the lowest jitter among the RV series evaluated. Based on these findings, we determine that the GND\_2.8 series represents the optimal dataset for our analyses. Consequently, we will focus exclusively on this time series in our subsequent work.
}

\begin{figure*}
    \centering
    \includegraphics[width=\textwidth]{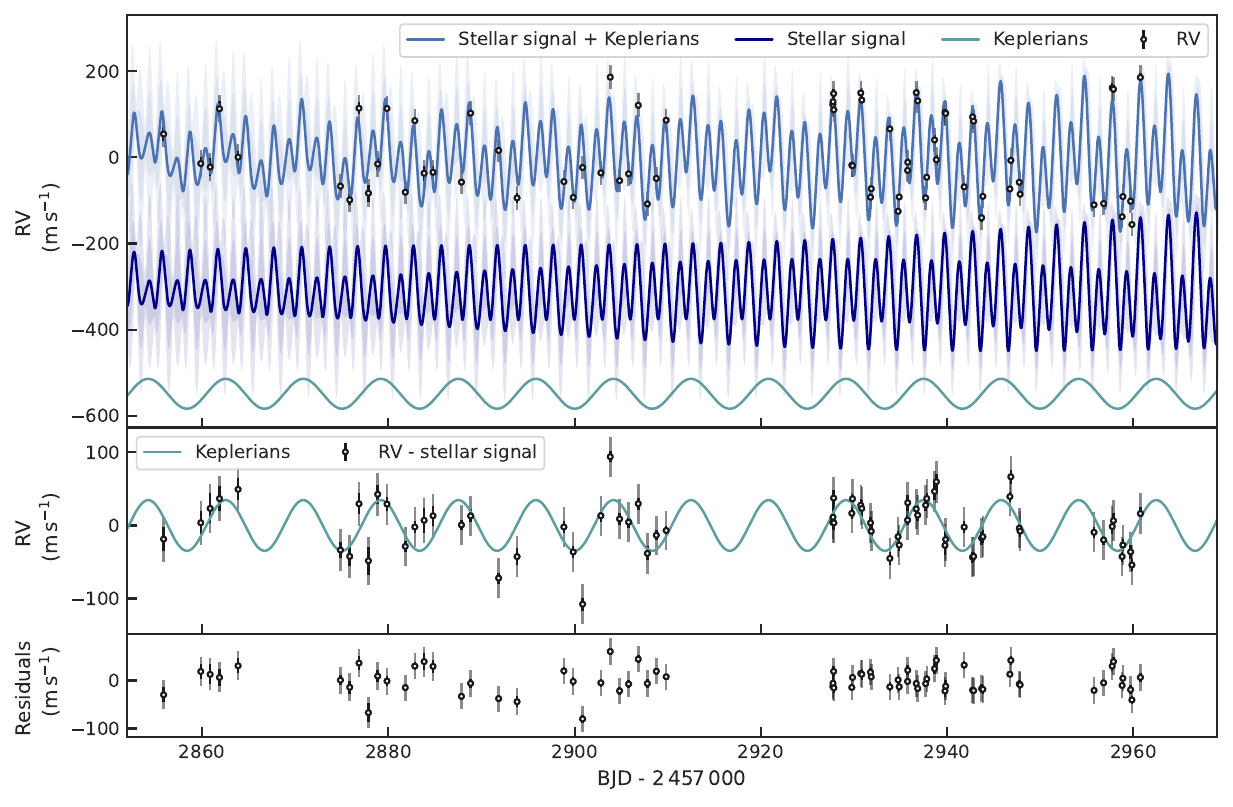}
    \caption{\target's RV time-series after being corrected by inferred offsets. \emph{Top panel:} RV data together with full, stellar and planetary signal inferred models; RV data with stellar signal model subtracted; and RV residuals.
    Measurements are shown with black circles, error bars, and a semi-transparent error bar extension accounting for the inferred jitter. 
    The solid lines show the inferred full model coming from our multi-GP, light-shaded areas showing the corresponding GP model's one and two-sigma credible intervals.
    We also show the inferred stellar (dark blue line) and planetary (light green line) recovered signals with an offset for better clarity.
    \emph{Middle panel:} RV data with stellar signal model subtracted and planetary model. 
    \emph{Bottom panel:} RV residuals.
    }
    \label{fig:timeseries}
\end{figure*}

\begin{figure}
    \centering
    \includegraphics[width=0.45\textwidth]{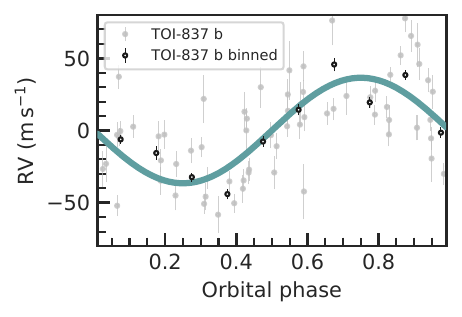}
    \caption{Phase-folded RV signal of \targetb\ following the subtraction of the systemic velocity and the stellar signal model. The solid line shows the inferred model. Nominal RV observations are shown as light grey points. Solid colourful points show binned data to 1/10 of the orbital phase. }
    \label{fig:rvphase}
\end{figure}

\subsection{Final joint model}
\label{sec:final}

Based on the analyses presented in this section, our final model for the photometric data of \target\ is the transit model described in Section~\ref{sec:transitanalysis}, together with the RV model \oscar{to the GND\_2.8 data} described in Sect.~\ref{sec:rvanalysis}. 
For completeness, we ran a final joint model of photometry and RVs to characterise \targetb.
The whole set of sampled parameters and priors are shown in Table~\ref{tab:pars}.

We also explored a solution where we allowed for an eccentricity. We inferred an orbital eccentricity of $0.10_{-0.05}^{+0.06}$. However, the model including a circular orbit is slightly preferred with $\Delta\,{\rm AIC} = 4$. This suggests that the orbit of \targetb\ is (close to) circular and the data we have is not enough to measure any small deviation from it.

Figure~\ref{fig:timeseries} shows the RV time series, while Fig.~\ref{fig:rvphase} shows the phase-folded Doppler signal.
Table~\ref{tab:pars} shows the inferred sampled parameters, defined as the median and 68.3\% credible interval of the posterior distribution.
Table~\ref{tab:pars} also shows the derived planetary and orbital parameters.

\section{Discussion}
\label{sec:discusion}

\subsection{RV detection tests}

We know a priori that \targetb\ exists and its nature is consistent with being planetary \citepalias{Bouma2020}. 
Therefore, we expect that there is a Doppler signal larger than zero that is consistent with the ephemeris of the transiting signal.
If we assumed that \targetb\ has the same properties as similar planets with its radius, we estimated a mass of $\sim 0.11$\,\mjup\ that would generate a Doppler signal of $\sim 10$\,\ms \citep{ChenKipping2017}.
Due to its youth ($\sim 35$\,Myr), we would expect a rather lower density object with a significantly smaller mass \citep[e.g.,][]{Baraffe2008,Fortney2007}.
Our detected Doppler semi-amplitude is \kb\ which translates to a mass of \mpb\ for \targetb. This mass together with the inferred radius of \rpb\ results in a planetary density of \denpb. 
This unexpected high density on young giant planets has been found previously \citep{Suarez2022}, but the RV detection of such planets are challenged by the community \citep[e.g.,][]{Blunt2023}. 
In this section, we test the reliability of our detection of the induced Doppler signal on \target.

\subsubsection{Planet signal in periodograms}

As shown in Section~\ref{sec:periodograms} and Figure~\ref{fig:periodiograms} there is no evidence of the planetary signal in RV GLS periodograms of \target. This is expected given that any planetary signal would be dwarfed by the relatively high stellar activity signal. 

We then took the residuals that we obtained from the no-planet onedimensional GP analyses presented in Sect.~\ref{sec:rvanalysis} \oscar{for the tree different RV time series and we applied a GLS periodograms that are shown in the bottom panel of Figure~\ref{fig:periodiograms}}.
We can see that the peaks that correspond to the stellar signal disappeared from the periodograms, indicating that the GP model was able to remove the intrinsic stellar signal in the RV time series. 
Furthermore, a significant peak appeared close to the orbital period of the planet. 
This result is unexpected. GPs have been proven to engulf planetary signals, especially when the latter ones are not included in the model \citep[e.g.,][]{Ahrer2021,Rajpaul2021}. 
We speculate that if a relatively strong coherent signals is present in our data set, it may survive a GP regression if it evolves with significantly different time scales than the stellar signal.
This assumption is consistent with the recovered Doppler signal of \targetb.

\begin{figure}
    \centering
    \includegraphics[width=0.49\textwidth]{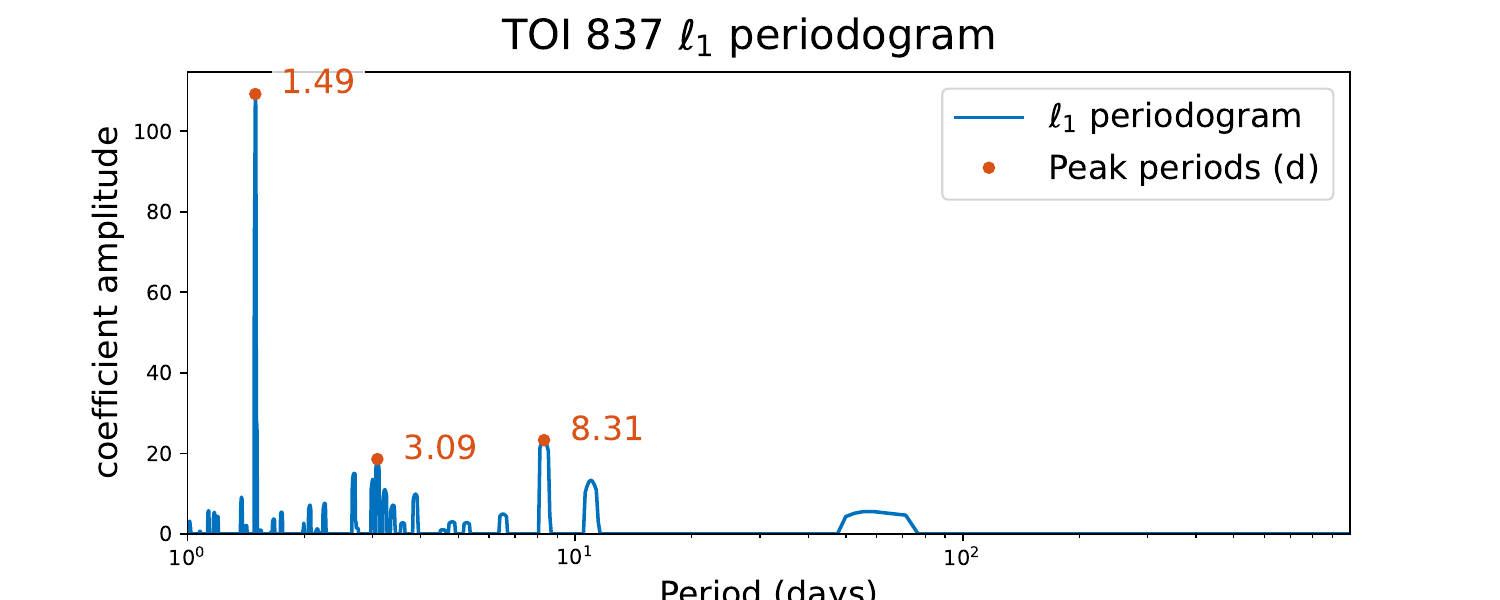}
    \caption{$\ell$1-periodogram for \target\ RVs.}
    \label{fig:l1p}
\end{figure}

To explore further the existence of the planetary signal in the periodogram, we ran a $\ell 1$ periodogram \citep{Hara2017} on the raw \oscar{GND\_2.8} RV time series.
Briefly, the $\ell$1-periodogram is a variation of the Lomb-Scargle periodogram methodology that has reduced sensitivity towards outliers and non-Gaussian noise.
Figure~\ref{fig:l1p} shows the $\ell$1-periodogram applied to the raw RV time-series of \target.
The three most significant peak happens at 1.5, 8.31 and 3.09\,d that correspond to the first harmonic of the rotation period of the star, the planet orbit, and the rotation period of the star, respectively. 
This suggests that there is a strong deep-seated signal that coincides with the orbital period of \targetb.
This result supports our previous speculation that there is evidence of strong coherent signal in the RVs that is consistent with our \targetb's recovered Doppler signal.

\subsubsection{Cross-validation}

\begin{figure*}
    \centering
    \includegraphics[width=0.98\textwidth]{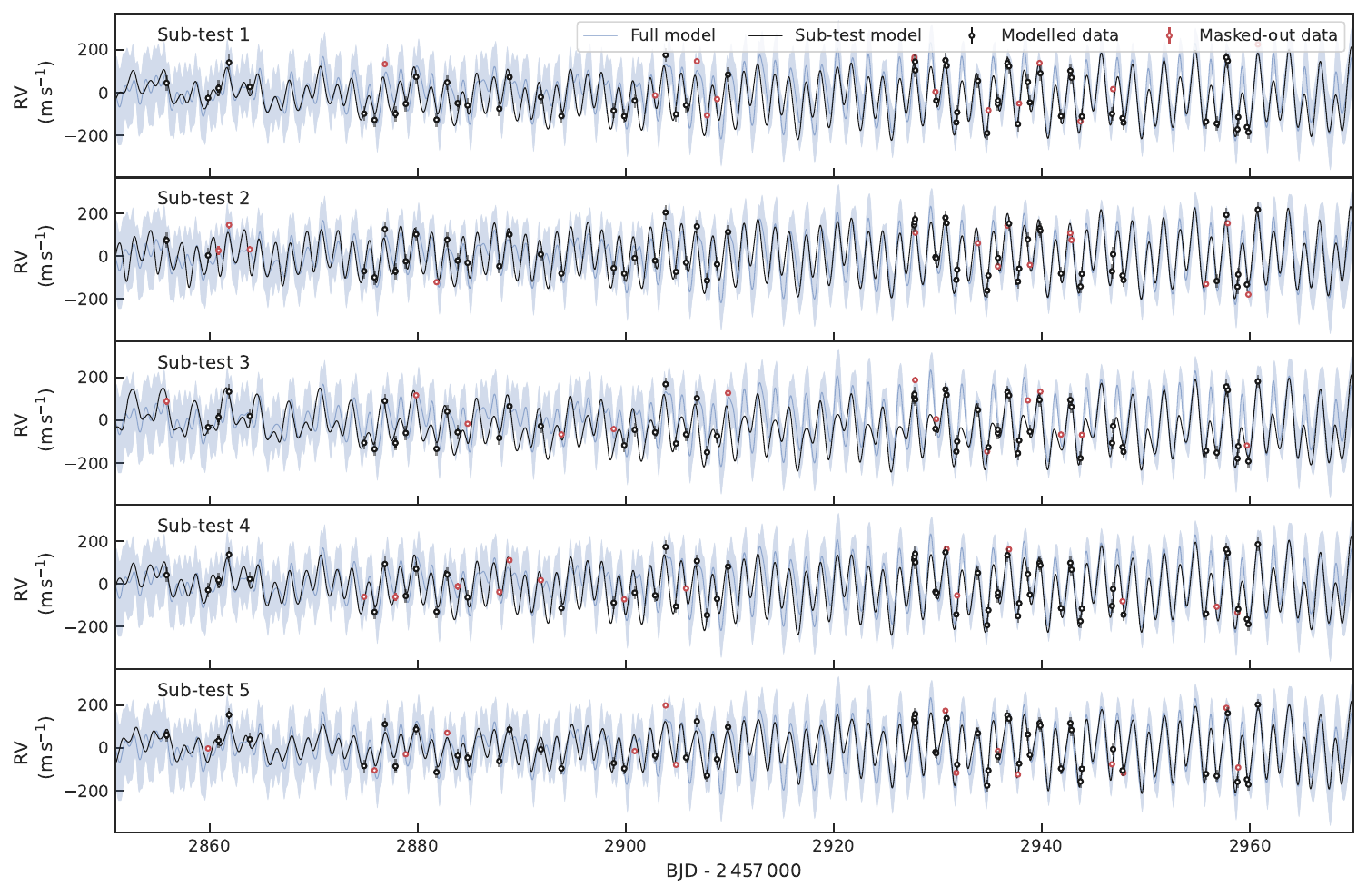}\\
    \includegraphics[width=0.98\textwidth]{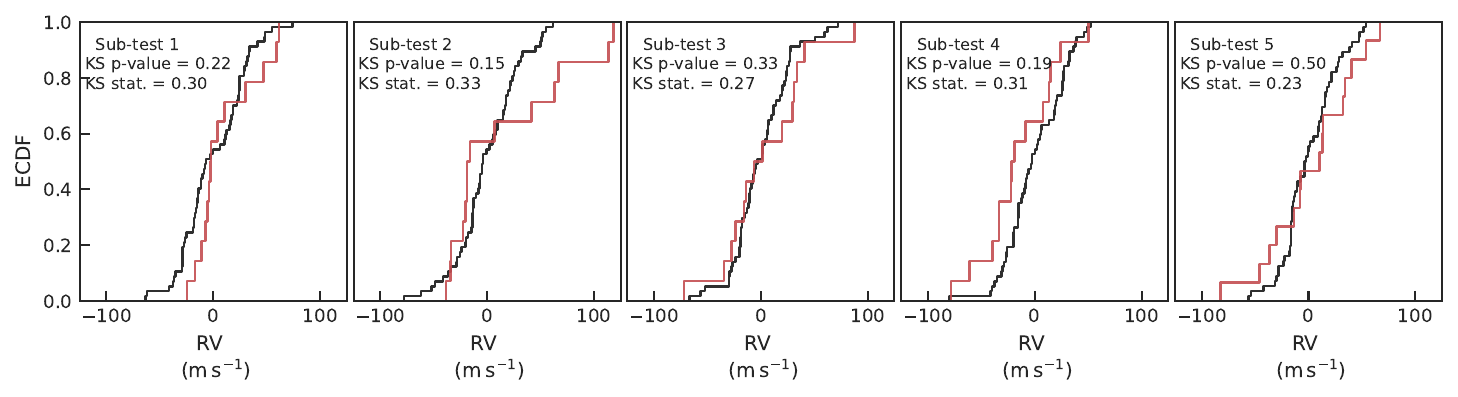}
    \caption{\emph{Top panel}: Cross-validation results with RV data sub-sets. Dark circles represent modelled data, red circles indicate masked-out data. Black solid line shows the recovered model for each sub-test. We also show the full model and $3-\sigma$ confidence interval of the full model recovered from the whole data set (same inferred model as Figure~\ref{fig:timeseries}) with a blue line and light shaded areas, respectively.
    \emph{Bottom panel}: Empirical Cumulative Distribution Functions (ECDF) for the residuals of the modelled (black) and masked-out (red) data for all sub-tests. Each sub-plot shows the corresponding KS statistics.
    }
    \label{fig:timeseries_cv}
\end{figure*}

The next test we perform is cross-validation to check for overfitting as suggested by \citet{Blunt2023}. 
To do so, we employ a data partitioning strategy similar to the principles of k-fold cross-validation \citep[see e.g.,][]{Fazekas2024} to systematically divide our dataset into distinct subsets. 
Specifically, we first randomised the dataset to ensure that each partition is representative of the overall data distribution. Subsequently, we segmented the dataset into five partitions. 
In each iteration, we use four out of these five partitions (constituting approximately 80\% of the data) for the modelling, while the remaining partition (approximately 20\% of the data) is masked out. 
This approach guarantees that each unique subset serves as the test set exactly once, eliminating any potential overlap between masked-out data across all iterations. 
Figure~\ref{fig:timeseries_cv} shows the five different sub-samples created with this approach.

We then perform a RV analysis (accounting for the planet signal) as the one presented in Sect.~\ref{sec:rvanalysis} for each sub-sample, we call each one of these runs as sub-test $i$, where $i$ runs between 1 and 5.
Figure~\ref{fig:timeseries_cv} shows the data and recovered model for all cases.
The recovered Doppler semi-amplitude for each case is fully consistent with the value reported in Sect.~\ref{sec:rvanalysis}, but with slightly larger error bars, as expected due to the smaller number of data modelled.
We can see that, by eye, the masked-out points fall well within the limits of the inferred models in each case, suggesting that the predictive model in each case can explain the masked-out observations \citep[see also discussion in][]{Luque2023}.
We can also see that the predictive model for each sub-set (black lines) falls well within the confidence intervals of the predictive distribution of the model of the full data set (blue lines and shaded region).
These results suggest that our model does not suffer overfitting.

To assess further the reliability of our cross-validation, we perform a Two-sample Kolmogorov-Smirnov \citep[KS;][]{Kolmogorov1933,Smirnov1948} test.
Briefly, the KS test is a non-parametric statistical test that is utilised to validate the hypothesis that two subsets of data originate from the same continuous distribution. 
This test compares the empirical cumulative distribution functions (ECDFs) of two samples without making any assumptions about the underlying distribution of data.
The maximum absolute difference between the ECDFs of the two samples, known as the KS statistic, serves as the basis for evaluating the null hypothesis that the two samples are drawn from the same distribution. A low KS statistic, accompanied by a high p-value, indicates a failure to reject the null hypothesis, suggesting that the differences between the two samples could be attributed to random variation, and hence, they can be considered as coming from the same distribution.

We apply the KS test to each one of our sub-tests in which we compare the residuals of the modelled data with the residuals of the masked-out data (i.e., masked-out points minus predictive distribution at their respective times). 
If the KS test proves that both samples are explained from the same distribution, then we can conclude that the predictive model explains well the masked-out data. This implies that the inferred model has the ability to explain unseen data, thus, it does not overfit.
Figure~\ref{fig:timeseries_cv} shows the ECDF for the modelled and masked-out data for each one of the sub-tests.
The p-values of the five sub-tests are significantly larger than 0.05, therefore, we can conclude that both samples are consistent with being described by the same underlying distribution.
These results give us confidence that our RV model for \target\ does not suffer from overfitting.

\subsubsection{Injection tests}

Previous research indicates that complex models can generate false planet-like RV time series, particularly in stars with high activity levels \citep[e.g.,][]{Rajpaul2016}. 
To check if this is our case, we performed injection tests similar to the ones presented in \citet{Barragan2019,Zicher2022}. 
We used \citlalatonac\ \citep{pyaneti2} to simulate synthetic RV time series at the same time-stamps as our HARPS data.
We first took the predictive distribution for the stellar signal that we obtained for \target\ in Sect.~\ref{sec:rvanalysis}.
We then added correlated noise using a squared exponential kernel with a length scale of one day, and the same amplitude as the jitter term obtained from the real data to simulate the red noise in our data.
To finish, we also added white noise for each synthetic observation according to the nominal measurement uncertainty of each HARPS datum.
We created 100 different realisations of stellar-like signals.
For each stellar-like signal, we created 3 different types of RV time series injecting signals with Doppler semi-amplitudes of 0 (assuming there is no planet), 10 (expected signal of this planet if it were inflated), and \oscar{35}\,\ms (the recovered signal). 
This leads to a total of 300 synthetic RV time series with similar noise properties and the same time sampling as the real data. 

\begin{figure}
    \centering
    \includegraphics[width=0.48\textwidth]{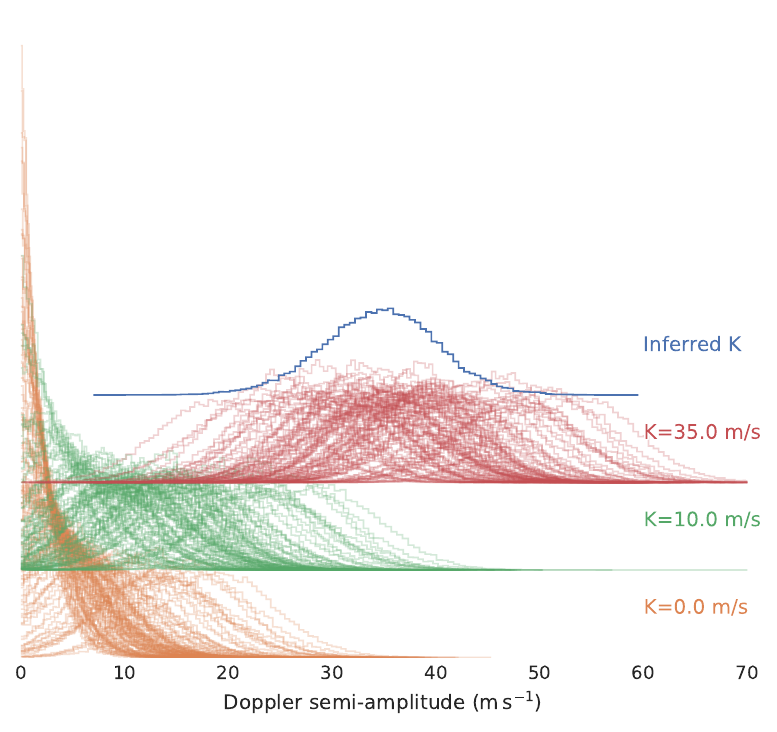}
    \caption{Doppler semi-amplitude distributions from injection tests. The blue line shows results from original data; other coloured lines from synthetic data. Labels indicate injected values.}
    \label{fig:injection}
\end{figure}

We model each synthetic dataset using a one-planet model and onedimensional GP configuration as described in Sect.~\ref{sec:rvanalysis}. 
We plot the recovered posterior distribution of the Doppler semi-amplitude of the planet for all the runs in Figure~\ref{fig:injection}.

We first check the percentage of the simulations in which we can claim a detection.
For this purpose, we consider that a detection has occurred if the median of the posterior is larger than 3-$\sigma$, where $\sigma$ is half the interval between the $16.5^{\rm th}$ and $83.5^{\rm th}$ percentiles.
We found that we can claim a detection of \oscar{3\%, 28\%, and 100\%} for the synthetic time series with injected signals of 0, 10, and \oscar{35}\,\ms, respectively.
We then revise which percentage of these detections is consistent with our detection in the real data set. We check what fraction of the time the recovered semi-amplitude in the synthetic time series is within 1-$\sigma$ of the recovered posterior distribution from the real data set.
We found that this condition is filled in \oscar{0\%, 1\%, and 62\%} of the cases for the 0, 10, and \oscar{35}\,\ms\ time series, respectively. 
These results suggest that if the real Doppler signal presented in our RV time-series was 10\,\ms\ we would detect the \oscar{35}\,\ms\ signal with a probability of \oscar{1\%}.
The \oscar{62}\% recovered in the \oscar{35}\,\ms\ case is consistent with the expected value taking into account Poisson counting errors.
\newline

For the remainder of the paper, we will assume that our mass and radius measurements of \targetb\ are reliable. Below we describe the physical implications of this for the planet's properties.

\subsection{TOI-837b's properties}

\begin{figure*}
    \centering
    \includegraphics[width=0.98\textwidth]{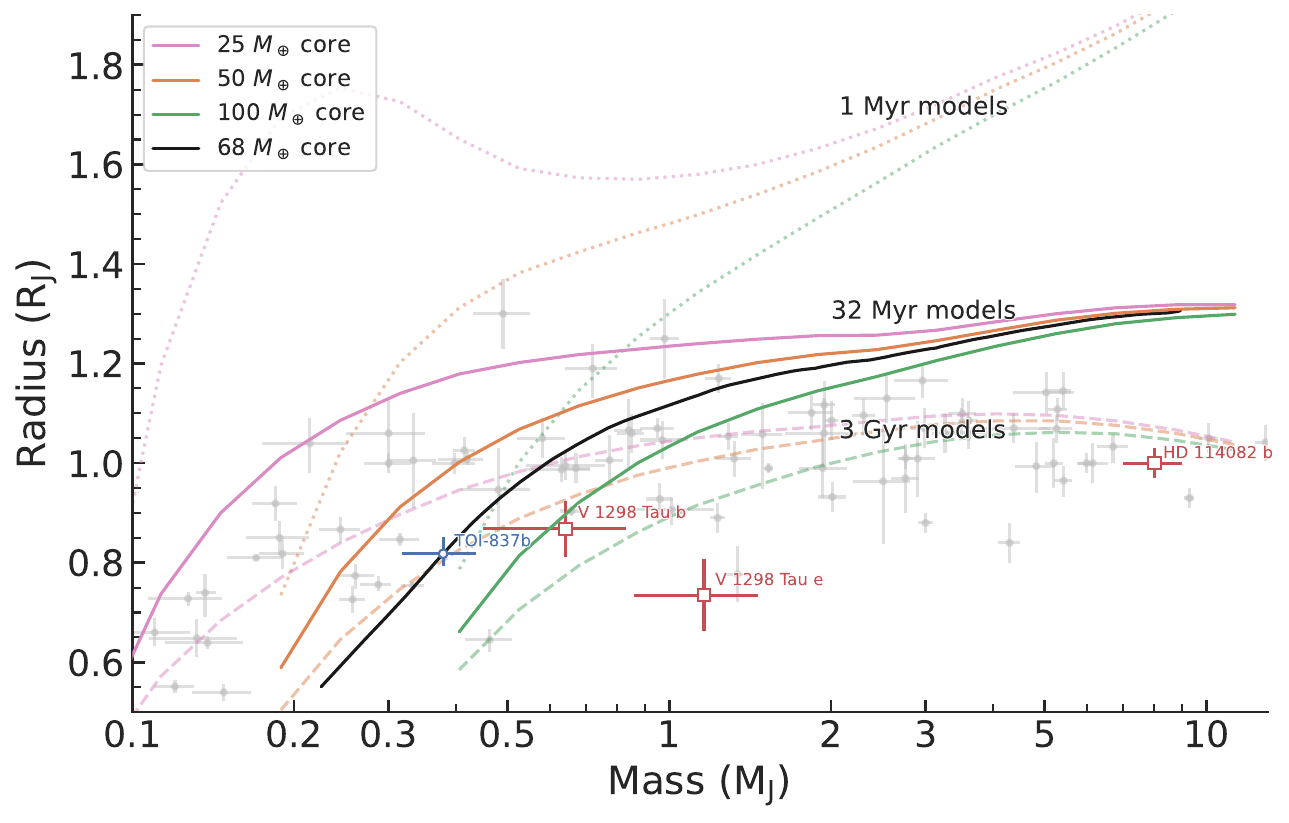}
    \caption{Mass vs Radius Diagram. Grey data points with error bars show exoplanets with mass and radius measurements with a precision of 30\% or better and orbital semi-major axis between 0.08 and 1\,AU. Data taken from the NASA Exoplanet Archive as of March 2024 \citep[\url{https://exoplanetarchive.ipac.caltech.edu/};][]{NASAexoplanet}. 
    The location of \targetb\ is highlighted by a blue circle. Solid lines depict theoretical models predicting the mass-radius relationship for planets of varying core masses, assuming an age of 32 Myr and a distance of 0.1 AU from a Sun-like star, based on \citet{Fortney2007}.
    Dotted and dashed lines extend these predictions to planets aged 1 Myr and 3 Gyr, respectively.
    Black line shows the interpolated model for a 32\,Myr old and \oscar{68}\,\mearth\ core.
    \oscar{
    We also highlighted other giant exoplanets younger than 100\,Myr with red squares.}
    The methodology for generating this diagram follows the approach outlined in \citet{Barragan2018b}.
    }
    \label{fig:mr}
\end{figure*}

With a mass of \mpb\ and radius of \rpb, \targetb's position in the mass-radius diagram is highlighted in Fig.~\ref{fig:mr}.
Figure~\ref{fig:mr} also shows a mass-radius diagram for giant exoplanets ($0.5 < R_{\rm J} < 2\,$\rjup\ and $0.1 < M_{\rm p} < 13$\,\mjup) detected with a precision better than 30\% in radius and mass. 
Considering that the evolution of planets is influenced by their distance from their host star, we only present planets that orbit within a semi-major axis range of 0.08 to 1 AU. This selection is informed by the semi-major axis of \targetb, $\sim 0.09$\,AU, and is supported by models that forecast a comparable evolution for planets situated within this range of semi-major axes \citep[see][]{Fortney2007}. 
We also over plot the \citet{Fortney2007}'s models with similar parameters to those of \targetb. This corresponds to models for planets around a solar-like star, with an age of 32\,Myr, at a distance of 0.1 AU with cores of 25 (pink), 50 (orange) and 100 (green)\,\mearth. 
We performed a radial basis function interpolation on the aforementioned \citet{Fortney2007}'s models to find the core mass that best describes the mass and radius of \targetb. 
We found that the position and age of \targetb\ would be consistent with a young planet with a core mass of  \oscar{$68\,M_\oplus$}, assuming a 50-50\% ice-silicate core (black line in Fig.~\ref{fig:mr}). 
This core mass corresponds to $\sim 60\%$ of \targetb\ total mass of $\sim 125\,M_\oplus$.

From figure~\ref{fig:mr}, we can see that if well there are planets that fall close to \targetb\ in the mass vs radius diagram, they all are older than 1\,Gyr and they should be compared with older planet's models. 
This can be done from figure~\ref{fig:mr} where we also show the \citet{Fortney2007}'s models for planets aged 3\,Gyr with orbital semi-major axis of 0.1~\,AU.
The first thing we see is that the older counterparts of \targetb\ are consistent with planets with cores $\lesssim 50\,M_\oplus$. If \targetb\ itself was older, we would estimate a significantly smaller core of $\sim 50\,M_\oplus$, our estimation of $\sim 70\,M_\oplus$ is a pure consequence of the expected inflated state, hence larger radius, given its youth.
In this line of thought, \targetb\ stands out as the only planet with a semi-major axis of $\sim 0.1$\,AU consistent with a $\sim 70\,M_\oplus$ core.

We then relaxed our assumptions and proceed to compare \targetb\ with planets with similar mass and radius, but without constrain on their semi-major axes.
We found that K2-60\,b \citep{Eigmuller2017}, HD~149026\,b \citep{Sato2005}, and TOI-1194\,b \citep{Wang2023} are denser than \targetb\ and are consistent with core masses between $50-100$\mearth\ despite being significantly older ($ > 1$\,Gyr). 
This suggests that planets with relatively similar masses to \targetb\ and large cores can exist.
The question now is what physical mechanisms can create such massive cores. 
\citet{Johnson2012} theorised that a star's high metallicity might reflect a primordial circumstellar disc with elevated dust content that could potentially enhance core formation.
Both,  HD~149026\,b and TOI-1194\,b have super-solar atmospheric metallicities that are consistent with this model. 
In fact, \citet{Bean2023} observed a significant atmospheric metal enrichment on the atmosphere of  HD~149026\,b that is explained by a bulk heavy element abundance of 66\% of the planetary mass.
However, the solar metallicity of \target\ and K2-60 does not support this hypothesis, suggesting that other factors may also play a critical role in the formation of massive cores.
\citet{Boley2016} proposed a theory where massive cores, with masses exceeding $20 M_{\oplus}$, could result from the merging of tightly packed orbiting inner planets that formed during the initial phases of the circumstellar disc's evolution. 

Another possibility is that \targetb\ has a less massive core, but the models are biased towards larger radii for young planets.
Core-accretion models of planetary evolution predict planets of
$<100\,$Myr to be at the early stages of their contraction phase, showing very large radii and low densities \citep[][see also 1\,Myr models in Fig.~\ref{fig:mr}]{Baraffe2008,Fortney2007}.
However, \citet{Fortney2007}'s models do not include a formation mechanism and can be arbitrarily large and hot at very young ages \citep[see][]{Marley2007}. This can lead to giant planet's models with radii several tenths
of a Jupiter radius larger than one computed taking into account formation mechanisms. This would lead as a consequence to estimate more massive cores for a given planet's mass and radius.
For the remainder of the paper we will assume that \targetb\ can be described by \citet{Fortney2007}'s models assuming a core of $70$\,\mearth.

\oscar{
It is also worth to compare \targetb\ with the properties of other exoplanets that are younger than 100\,Myr. According to the NASA exoplanet archive \citep[][]{NASAexoplanet}, there are five such exoplanets documented with precise measurements of age, mass, and radius. These planets include AU Mic b and c \citep[22\,Myr,][]{Plavchan2020,Martioli2021,Zicher2022,Donati2023}, V1298 Tau b and e \citep[23\,Myr,][]{Suarez2022,Finociety2023}, and HD 114082\,b \citep[15\,Myr,][]{Zakhozhay2022}.
The planets orbiting AU Mic have properties that suggest they are inflated when compared to exoplanets orbiting older stars. This aligns with existing theoretical models that predict younger planets might exhibit such inflated characteristics due to active and intense processes in their early developmental stages. Although these planets are smaller than Neptune, with radii smaller than 4\,$R_\oplus$, they showcase a distinct set of physical processes which might differ significantly from those observed in larger exoplanets like \targetb.
Moreover, the exoplanets V1298 Tau\,b and e, along with HD 114082\,b, are consistent with giant planets, as illustrated in Fig.~\ref{fig:mr}. Interestingly, these planets are denser than what current theoretical models predict for their age and type \citep[see discussions in][]{Suarez2022,Finociety2023,Zakhozhay2022}. This tendency suggests that existing theoretical models and/or mass/radius young exoplanet measurements require further dedicated studies. 
In case our measurement methods are correct, this implies a reconsideration of planetary formation and evolution theories.}

\subsection{Atmospheric characterisation perspectives}

\begin{figure*}
    \centering
    \includegraphics[width=0.9\textwidth]{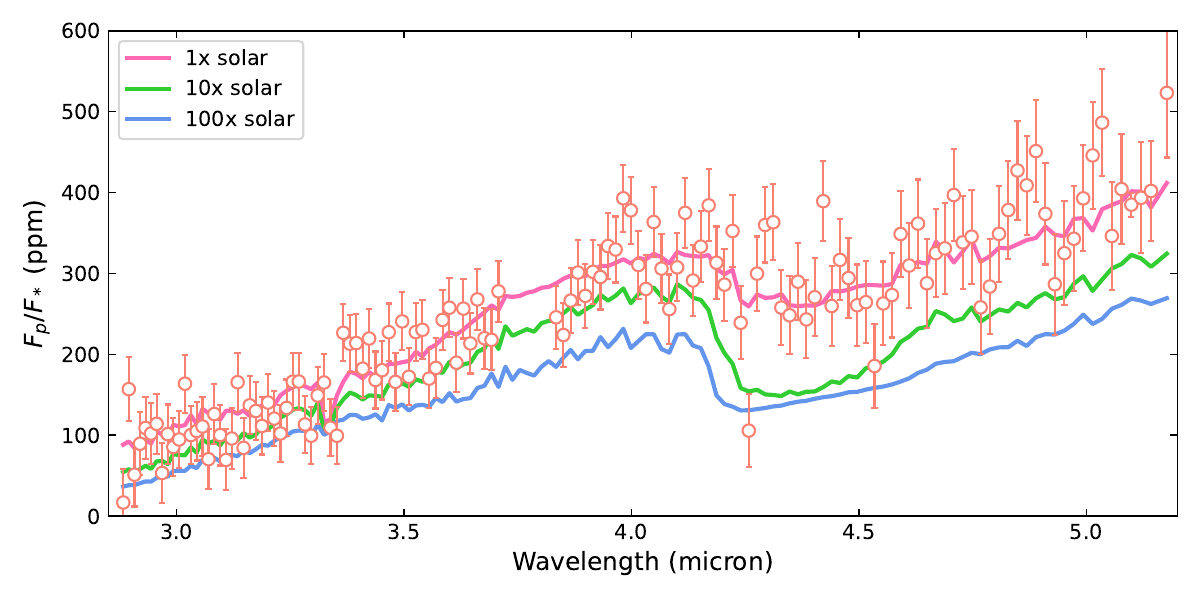}
    \caption{The predicted emission spectrum of \targetb. The red data points show the spectrum attainable with three eclipse observations, using JWST/NIRSpec G395H, assuming a solar composition in both metallicity and C/O ratio. For comparison, the forward-modelled spectra assuming $10\times$ and $100
    \times$ metallicity are also shown (with solar C/O).}
    \label{fig:jwst-nirspec}
\end{figure*}


Atmospheric observations could play a crucial role in differentiating among various potential formation mechanisms for \targetb. According to the mass-metallicity relationship empirically established by \citet{thorngren2016}, our mass measurement of \targetb\ suggests a predicted metal fraction ratio between the planet and its host star of $Z_\mathrm{planet}/Z_\mathrm{star}=14.7\pm 2.5$. 
With \target's stellar metallicity measured at $\mathrm{[Fe/H]}=0.01\pm0.04$, this translates into an estimated bulk metallicity fraction for \targetb\ of $0.21\pm0.04$.
\citet{Bean2023} presents a relationship between atmospheric and bulk metallicity fractions. If we assume that \targetb\ follows this trend, we could expect its atmospheric metallicity to surpass solar values by a factor of $5-8$. 
This significant enhancement in atmospheric metallicity would not only support the substantial core mass inferred from our observations but also provide insight into the planet's formation conditions and subsequent evolutionary history.

To predict the observability of the atmosphere of \targetb\ with \jwst, we leverage packages \textsc{picaso}\footnote{\url{https://natashabatalha.github.io/picaso/}} and \textsc{pandexo}\footnote{\url{https://natashabatalha.github.io/PandExo/index.html}}.
We forward model the emission spectrum of \targetb\ using \textsc{picaso} \citep{batalha2019}, based on the parameters derived in this work.
We make use of the integrated 1D climate modelling for the temperature-pressure (TP) profile and chemical abundance solution \citep{mukherjee2023}, including all opacities available in \textsc{picaso}.
The TP profile includes a convective zone only in the deepest atmospheric layers.
We then use \textsc{pandexo} to estimate the signal-to-noise of the dayside emission signal of \targetb\ with \textit{JWST/NIRSpec G395H}; the predicted observations using three eclipses are shown in Fig.~\ref{fig:jwst-nirspec}, assuming solar metallicity and C/O. 
At least two eclipse observations are needed to distinguish between a $1\times$ and $10\times$ solar metallicity scenario.
Using the chi-square statistic for hypothesis testing, we would expect to distinguish between these cases to a significance of $11.1\sigma$ with two eclipse observations (totalling 10.01\,hrs), rising to $17.1\sigma$ with three eclipses (Fig.~\ref{fig:jwst-nirspec}).
Constraining the atmospheric metallicity will help to break the degeneracy with interior composition and the implications for formation.

\section{Conclusions}
\label{sec:conclusions}

Our investigation into the \target\ system and its intriguing companion, \targetb, unveils a young Saturn-sized exoplanet that defies conventional expectations with its unexpected massive core. Our exhaustive analysis of data from \tess\, ground-based observations, and HARPS spectroscopic enabled us to determine \targetb's radius at \rpb\ and mass at \mpb, translating to a density of \denpb. This density together with its age and distance to the star suggest a core mass of approximately 70 $M_{\oplus}$, accounting for 60\% of the planet's total mass. Such a substantial core within a relatively young planetary body presents a challenging scenario for current models of planet formation and core accretion, especially due to the relatively low stellar metallicity.

The unique characteristics of \targetb\ underscore the urgency for advanced atmospheric characterisation. Eclispe observations with \jwst\ could to offer unparalleled insights into the composition of \targetb. 
A measurement of the planetary atmospheric bulk metal fraction will potentially elucidate the true nature of its significant core. Such future studies are crucial for breaking the current degeneracies in planet composition models and could revolutionise our understanding of planetary formation.

We also leave open the possibility that our RV detection could not be accurate. Despite the tailored campaign with a high cadence of observations, the apparent fast evolution of the strong stellar signal could generate biased measurements of the Doppler semi-amplitude. We showed with several statistical tests that our RV planetary detection is robust, nevertheless, further observations could help us to test this further. For example, tailored near Infrared RV campaigns, where stellar activity is less significant, with instruments such as the Near Infra Red Planet Searcher (NIRPS) could help to assert our detection in the optical.

\section*{Acknowledgements}
{
This publication is part of a project that has received funding from the European Research Council (ERC) under the European Union’s Horizon 2020 research and innovation programme (Grant agreement No. 865624).
This work made use of \texttt{numpy} \citep[][]{numpy}, \texttt{matplotlib} \citep[][]{matplotlib}, and \texttt{pandas} \citep{pandas} libraries.
This work made use of Astropy:\footnote{\url{http://www.astropy.org}} a community-developed core Python package and an ecosystem of tools and resources for astronomy \citep{astropy1, astropy2,astropy3}. 
This publication made use of data products from the Wide-field
Infrared Survey Explorer, which is a joint project of the University of
California, Los Angeles, and the Jet Propulsion Laboratory/California
Institute of Technology, funded by the National Aeronautics and Space
Administration.
\oscar{We thank the anonymous referee for their comments/suggestions that helped to improve the quality of this manuscript.}
AVF acknowledges the support of the IOP through the Bell Burnell Graduate Scholarship Fund.
\oscar{
OB thanks Luke Bouma for the insights on the up-to-date literature on the TOI-837 system.
}
OB acknoweledges that the random seed used for the injection simulations was 060196.
}

\section*{Data Availability}

The codes used in this manuscript are freely available at \url{https://github.com/oscaribv}.
The spectroscopic measurements that appear in Table~\ref{tab:harps} are available as supplementary material in the online version of this manuscript.
All \tess\ data are available via the MAST archive. Ground-based     photometry are available on the online version of \citet{Bouma2020}.

%
\bibliographystyle{mnras} 
\bibliography{refs} 
%

\begin{appendix}

\section{Multidimensional GP analysis}
\label{sec:appendix}

\oscar{
In section \ref{sec:stellarsignal} we performed an analysis of different activity indicator time series \citep[as in][]{Barragan2023}. We found that they do not contain significant information about the stellar signal, more than the period. Therefore we conclude that they do not bring enough information that can help us to constrain better the shape of the stellar signal in the RV time series within a multidimensional GP framework.
}

\oscar{
As an educational test, we perform a two-dimensional GP analysis between the RV and FWHM time series. We follow the framework of \citet{Rajpaul2015} and \citet{pyaneti2} where we assume that RVs and FWHM are modelled as
\begin{equation}
\begin{aligned}
 \mathrm{RV} &=&  A_{\rm RV} G(t) &+ B_{\rm RV} \dot{G}(t) \\
 \mathrm{FWHM}& = & A_{\rm FWHM} G(t)& \\
\end{aligned}
\label{eq:gps}
\end{equation}
\noindent
where the variables $A_{\rm RV}$, $B_{\rm RV}$, and $A_{\rm FWHM}$ serve as free parameters that connect the individual time series with $G(t)$ and $\dot{G}(t)$. Here, $G(t)$ is considered a latent variable, meaning it is not directly observed. This variable can be broadly understood as representing the portion of the visible stellar disc that is covered by active regions over time.
}

\oscar{
We performed a multidimensional GP repression using \pyaneti\ and the same MCMC set-up described in Sect.~\ref{sec:transitanalysis}. We model the GND\_2.8 RVs together with the FWHM.  We use the QP kernel given in eq.~\eqref{eq:gamma} to construct the covariance matrix. The mean function for the FWHM was assumed to be an offset while for the RV was including a Keplerian following the guidelines presented in Sect.~\ref{sec:rvanalysis}. All priors follow the same guidelines presented in Table~\ref{tab:pars}.}

\begin{figure*}
    \centering
    \includegraphics[width=0.99\textwidth]{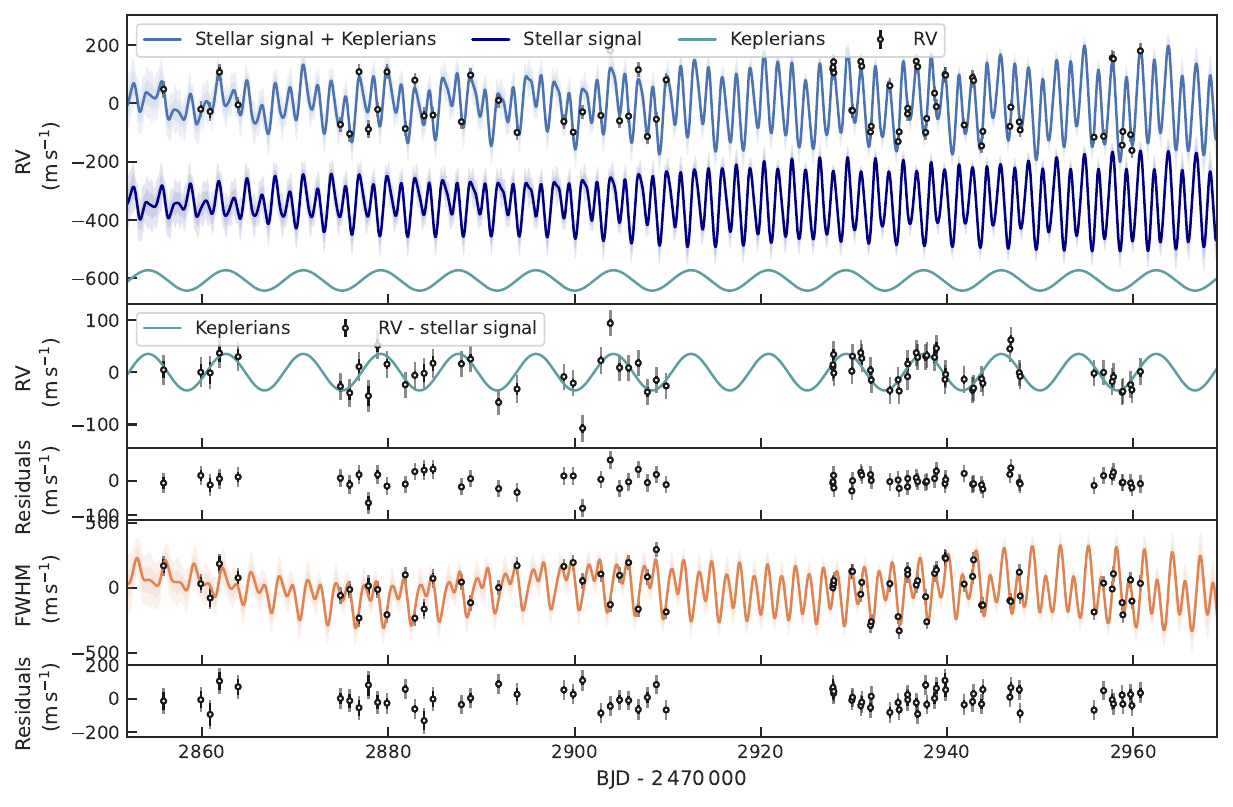}
    \caption{\oscar{ \target's RV and FWHM time-series after being corrected by inferred offsets. The plot shows (from top to bottom): RV data together with full, stellar and planetary signal inferred models; RV data with stellar signal model subtracted; RV residuals; FWHM data together with inferred stellar model; and FWHM residuals.
    Measurements are shown with black circles, error bars, and a semi-transparent error bar extension accounting for the inferred jitter. 
    The solid lines show the inferred full model coming from our multi-GP, light-shaded areas showing the corresponding GP model's one and two-sigma credible intervals.
    For the RV time-series (top panel) we also show the inferred stellar (dark blue line) and planetary (light green line) recovered signals with an offset for better clarity.}}
    \label{fig:mgp}
\end{figure*}

\oscar{
The recovered GP hyper-parameters are \lbe$=13.9_{-3.4}^{+4.1}$\,d, \lbp$=0.46_{-0.08}^{+0.11}$ and \pgp$=3.01\pm 0.02$\,d. The period is fully consistent with the results presented in Sect.~\ref{sec:rvanalysis}, as expected. We can see that the \lbp\ is relatively larger than the value reported in Table~\ref{tab:pars}. This is also expected given that in the multidimensional GP analysis, the high harmonic complexity in the RV time series is absorbed by the derivative term, $\dot{G}$, used to model the RVs \citep[see discussions in][]{pyaneti2,Barragan2022,Barragan2023}. 
On the other side, \lbp\ is also larger than the value recovered from the FWHM onedimensional analysis (see Table~\ref{tab:1gphp}).
This is not expected given that in both analysis we would expect that the FWHM constrain a similar shape and behaviour of the underlying $G(t)$ process.
The relatively high harmonic complexity found in Sect.~\ref{sec:ssactivityindicators} is likely caused by an overfitting that makes the model of the FWHM to vary faster than the sampling of the HARPS data. The lower harmonic complexity found in this case can be attributed to the RV time series setting a constrain on the shape of $G(t)$. In Figure~\ref{fig:mgp} we can see how the FWHM time series looks better constrained that the signal recovered in Figure~\ref{fig:timeseries1dgp}.
We also obtain a \lbe$=13.9_{-3.4}^{+4.1}$\,d that is smaller than the value obtained in the onedimensional RV analysis but similar to the value found in the FWHM analysis. This suggests that in this case, the FWHM constrain the evolutionary timescale. This has sense given that it is expected that a shorter timescale will dominate the evolutionary parameter of two signals with different \lbe\ values.
From this analysis, we can deduct that the multidimensional GP approach is able to model the stellar signal in the RV and FWHM time series. In this case, the RV time series allows to guide the shape of the $G(t)$ that permits a better modelling of the stellar signal in the FWHM time series. We perform another run when we model the Gaussian fit RVs and we obtained fully consistent results.
}

\oscar{
Regarding the planetary signal, we recover $K = 35.0 \pm 5.4$\,\ms\ that translates into a mass of $0.391 \pm 0.064$\,\mjup.
These values are fully consistent and almost identical with the ones presented in Sect.~\ref{sec:rvanalysis} and Table~\ref{tab:pars}. 
For a fit using the Gaussian fit RVs we recover $K = 36.4 \pm 5.9$\,\ms\ that translates into a mass of $0.391 \pm 0.064$\,\mjup.
We note that these results do not bring any significant improvement respect to the onedimensional GP model presented in Sect.~\ref{sec:rvanalysis} \citep[see also][]{Barragan2023}.
This can be explained by the lack of information in the activity indicators. 
Therefore, we have chosen to adopt the more conservative onedimensional regression approach for the main paper. This decision is driven by our judgement regarding the use of an activity indicator that alone does not constrain the stellar signal. 
}


\end{appendix}

\bsp	
\label{lastpage}
\end{document}